\documentclass[aps, twocolumn, prx]{revtex4-1}
\usepackage[utf8]{inputenc}
\usepackage[english]{babel}
\usepackage{graphicx,epsfig}
\usepackage{color}
\usepackage{amsmath,bbm,amssymb, amsthm}
\usepackage{ulem}
\usepackage{numprint}

\begin{document}

\title{Quantum technologies need a Quantum Energy Initiative}

\author{Alexia Auff\`eves}
\affiliation{Universit\'e Grenoble Alpes, CNRS, Grenoble INP, Institut N\'eel, 38000 Grenoble, France }

\begin{abstract}
Quantum technologies are currently the object of high expectations from governments and private companies, as they hold the promise to shape safer and faster ways to extract, exchange and treat information. However, despite its major potential impact for industry and society, the question of their energetic footprint has remained in a blind spot of current deployment strategies. In this Perspective, I argue that quantum technologies must urgently plan for the creation and structuration of a transverse quantum energy initiative, connecting quantum thermodynamics, quantum information science, quantum physics and engineering. Such initiative is the only path towards energy efficient, sustainable quantum technologies, and to possibly bring out an energetic quantum advantage.
\end{abstract}

\maketitle
%\pacs{}

\section{Introduction}
Quantum technologies were born out of blue sky research lines investigated in the eighties. Their efficacy relies on the most counter-intuitive features of quantum physics, a.k.a. quantum coherence and entanglement, that are exploited to process information more efficiently, to communicate it more safely, and to extract it more precisely than in the classical realm. Owing to huge experimental efforts in academic labs and more recently startups and large industry vendors, we enter an era where the potential impact of quantum technologies on national and industrial sovereignty has become clear to decision makers, giving rise to strategic programs all around the world and important funds allocated to quantum research both in the public and the private sector.  Addressing the challenges raised by quantum technologies requires to structure new interdisciplinary approaches, putting in synergy the work of quantum information science, engineering, and quantum physics. The concept of ``quantum engineering" \cite{Deutsch_PRXQuantum} captures this desirable hybridization of fundamental research and technology developments, which combines creativity, intuition skills, potential for scalability and care for practical use cases. 

Oddly enough, the question of the resources consumed by quantum technologies -- especially energy -- has only discreetly started to enter into the landscape. It is all the more surprising as the energetic footprint of classical Information and Communication Technologies (ICT) reaches new highs, representing $11\%$ of the global electricity consumption in 2020 \cite{Puebla}. Meanwhile gains in efficiency have saturated, and cannot compensate for the increasing number of users and use-cases -- giving rise to specific research and development programs \cite{Gammaitoni_2015}. In this view, quantum technologies can appear as a potential way to escape the energetic dead end, but there is currently a lot of confusion on the claims. On the one hand, gains in complexity provided by the quantum logic are put forward to anticipate potential energy savings. In this spirit, first estimations have compared the power consumed by Google's Sycamore quantum processor and by the supercomputer solving the same problem, revealing 5 orders of magnitude difference and pointing towards a quantum advantage of energetic nature \cite{Arute}. On the other hand, on the fault tolerant quantum computing side, it is generally acknowledged that the overhead of physical qubits used in error correction will be an issue on the road to scalability. This blatant lack of consensus on the apparently simple question: ``Is there an energetic quantum advantage?" -- not only on the answer, but also on the methodology to address the question -- demonstrates that we have barely scratched the surface of the energetic dimension, and that there is an urgent need to address it by putting together a proper combination of expertises.

\begin{figure}[h!]
\begin{center}
\includegraphics[width=0.5\textwidth]{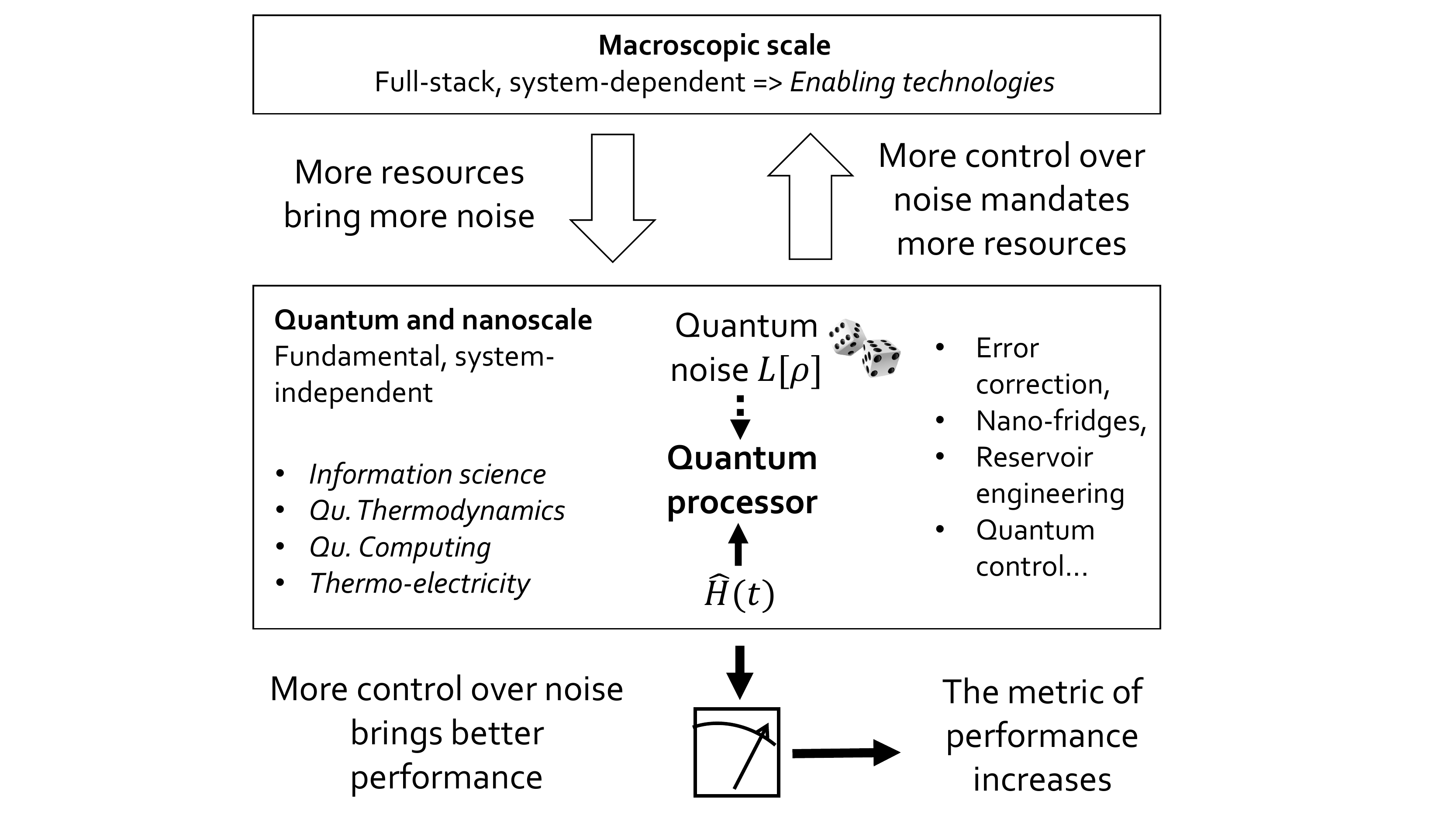}
\end{center}
\caption{Structure of the Quantum Energy Initiative (QEI) for quantum computing. The performance of a quantum computation is set at the fundamental level, and increases with the control over noise. More control over noise mandates more resources provided by the macroscopic level. These macroscopic resources bring more noise at the fundamental level. Optimized resource management relies on the alliance between different fields of expertise. This alliance defines the QEI called in the present Perspective.
\label{f:QEI}}
\end{figure}

In this Perspective, I argue that the deployment of quantum technologies urgently calls for the creation and structuration of a Quantum Energy Initiative (QEI, see Fig.\ref{f:QEI}), i.e. an interdisciplinary research line putting in synergy quantum thermodynamics, quantum physics, information science, and enabling technologies. Bridging this gap is more than timely. From a technological standpoint, only a tight connection between fundamental research and engineering can draw quantitative connections between the computing performances appearing at the quantum level and the energetic consumption at the macroscopic, full-stack level. Such considerations will directly impact technological choices and avoid dead-ends as quantum computers are being designed. Adopting such a transverse approach is also the only way to define energy based performance metrics, inspire smart and realistic resource optimization strategies, benchmark different qubits technologies or computing architectures, and rank the most energy efficient quantum processors. This global framework will shape the energetic future of information technologies, possibly bringing out a quantum energy advantage. 

I first present how the concepts of thermodynamics help mitigating energetic issues in the classical computing paradigm. In the second section, I propose a structure for the QEI. I finally draw a partial state of the art, that provides the first seeds of the initiative. My examples are drawn from quantum computing based on solid state qubits, but these ideas can be extended to any qubit type, allowing to benchmark them. They can also be developed for quantum simulation, quantum communication and quantum sensing. This whole Perspective intends to trigger a general awareness of the timeliness and relevance of energetic questions in quantum technologies. It also proposes objectives and roadmaps for future research calls and international collaborations.

\section{Thermodynamics of classical computing}

\subsection{Thermodynamic metrics} 
Thermodynamics was born in the nineteenth century from practical considerations, namely optimizing thermal machines like heat engines and fridges. It can be seen as the first physical theory that aimed to quantify and optimize the resource consumption of devices operating at a given level of performance -- allowing to define an efficiency as the ratio between the performance and the resource cost. As an example for heat engines, the resource (resp. the performance) equals the heat provided by the hot source (resp. the extracted work), giving rise to an efficiency with no physical dimension.  A major lesson of thermodynamics is the existence of fundamental bounds, that relate efficiency and thermodynamic time arrow: hitting them requires to operate in a thermodynamically reversible fashion. For instance, a heat engine efficiency cannot overcome Carnot bound. The bound is only reached if the device is run quasi-statically, i.e. if it delivers zero power, giving rise to a well-known trade-off between power and efficiency in heat engines \cite{CA_1975}. 

Along the twentieth century, the concepts of work, heat and irreversibility were extended to non-equilibrium systems where thermal fluctuations become predominant, defining the framework of stochastic thermodynamics \cite{Seifert2008,Peliti,Sekimoto}. Remarkably, irreversibility acquires a meaning at the level of single realizations of thermodynamic processes or ``stochastic trajectories", where it gets quantified through the stochastic entropy production. Stochastic entropy production gives rise to the celebrated fluctuation relations \cite{CJ_Book}, which connect the fluctuations of non-equilibrium quantities (e.g. the stochastic work) to the change of equilibrium ones (e.g. the free energy). Like nano-thermodynamics \cite{Hill}, stochastic thermodynamics provides deep insights into the thermodynamics of small systems, such as bits of information -- providing precious tools to quantify the energy cost of information processing. 

Very early in the history of computing, the founding fathers of information thermodynamics analyzed computers as thermal engines \cite{Bennett82}: a computer is a machine operating at finite temperature, inducing transformations on a ``data register" whose final state encodes the result of the computation. The register is a set of bits, i.e. physical systems characterized by two states denoted $0$ and $1$. A computing sequence typically consists of four steps, namely: (i) copy some external input state on the data register, initially prepared in a well defined reference state (ii) induce physical transformations on the data register, such that the result of the computation is encoded on its final state (iii) read the result (iv) reset the data register in its reference state. Step (iv) is logically irreversible: once the register has been reset, there is no way to retrieve the information it carried. It was argued by R. Landauer \cite{Landauer61} and C. Bennett \cite{Bennett82} that each such logically irreversible operation leads to some minimal amount of heat dissipation, namely $k_B T \log 2$ per erased bit. This bound is known as Landauer's bound: it scales like $10^{-21}$J at room temperature and is reached if the process is thermodynamically reversible. Importantly, logical irreversibility and thermodynamic reversibility peacefully coexist: logical irreversibility captures the erasure of the correlations between the external input state and the data register, a process which can be conducted at thermal equilibrium \cite{Sagawa14}. The cost of measurement and information erasure spent in steps (i) and (iv) have been shown to define the fundamental cost of information processing \cite{Sagawa-Ueda09}, solving the Maxwell demon paradox \cite{MD,Leff-Rex}.

Since the first computers implemented in the forties, huge progresses have been accomplished in energy management as it can be seen from their switching costs, i.e. the energy needed to turn $0$ to $1$ and vice-versa. While they scaled like a few mJ with vacuum-tubes, switching costs have continuously decreased to reach now $10^{-17}$J in CMOS technologies \cite{Gammaitoni_2015,EndMoore}. It is still four orders of magnitude above Landauer's bound. Such discrepancy is due to the lack of reversibility of CMOS based information technology, both thermodynamical and logical. On the one hand, one wants to compute ``as fast as possible", which is not consistent with thermodynamic reversibility:  one recovers the trade-off between speed and energy efficiency already encountered for heat engines. On the other hand, a classical gate-based computation involves NAND gates. CMOS based NAND gates input $2$ bits and output $1$, hence are logically irreversible. The search for better energy management has triggered explorations of alternative technologies and computing paradigms (see below). 

In classical computing, one cares about how fast one can process information: the computing performance is quantified by the number of floating point operations per second and expressed in FLOPs, usually with 32-bit numbers. Thus, a natural energy-based figure of merit is the ``Performance per Watt". This metric allows to benchmark different computing architectures or technologies and gave rise to the Green 500 ranking, that singles out the most energy efficient supercomputers. Expressed in FLOPs/W, the performance per Watt has the dimension of the inverse of an energy, and continuously increased over the years \cite{Oez}. The gains are not only due to hardware, but also to software improvements related to the computing architecture and the compilers. The best recorded performance is currently $40$ GFLOPs/W for a full system including cooling costs, and can exceed $100$ GFLOPs/W for individual Graphic Processor Units (GPUs) \cite{Desislavov}.

\subsection{Towards energy efficient computing}

To save energy, thermodynamics teaches us to compute in a thermodynamically reversible way. The basic principle can be grasped by modeling a bit of information as the two states of a particle in a double potential well \cite{Bennett82,Sergio, Gammaitoni_2015}. Information is stable when the potential barrier between the wells is large with respect to thermal fluctuations, but switching the bit state is expensive. Conversely, zero-energy switches and energy efficient erasures can be realized by slowly changing the size of the potential barrier. In this way, Landauer's bound was experimentally reached, e.g. with colloidal particles \cite{Sergio} and with adiabatic CMOS components, by shaping their voltage control \cite{ACMOS}. Interestingly, superconducting platforms such as Single Flux Quantum (SFQ), Rapid Single Flux Quantum, Adiabatic Quantum Flux Parametron (AQFP) are expected to provide gains both in clock frequency and power consumption \cite{Oez}.

To reduce even more the computing costs, a groundbreaking strategy is reversible computing. First proposed by Charles Bennett \cite{Bennett73}, reversible computing relies on the assumption that logically reversible operations can be implemented in a way that is energetically free. This mandates the use of reversible gates, i.e. that input and output the same number of bits and can be physically reversed. Once the computation is executed and the result is read, the idea is simply to ``rewind" the computation and ``uncompute" (i.e. inverse the step (ii) above), then uncopy (i.e. inverse the step (i)), such that the data register is brought back in its reference state. Such sequence allows to get rid of the erasure step, and brings the fundamental energy cost below Landauer's bound.

A pioneering example of reversible computer was the ballistic model proposed by Fredkin and Toffoli \cite{Fredkin_Toffoli}, in which billiard balls follow deterministic trajectories mapping computational paths. This model appeared to be very sensitive to errors, making thermal noise a foe \cite{Bennett82}. On the opposite, the model of reversible Brownian computing treats noise as a resource, letting probabilistic computing trajectories follow a minimal potential path towards the computation result \cite{Bennett82,Strasberg-Turing2015}. Reversible computing benefits from a revival of interest lately \cite{Frank18,Frank20}, as erasures below Landauer limit have been demonstrated on various platforms: Reversible Quantum Flux Parametron \cite{RQFP}, classical electronics \cite{Orlov_2012} and adiabatic CMOS (See e.g. in \cite{Snider_Book}). Probabilistic computing is another strategy to compute at low energy cost while exploiting thermal noise. Reminiscent of neural networks, it relies on probabilistic bits or ``p-bits" where the two logical states are separated by a low potential barrier \cite{Datta2019}. Brownian and probabilistic computing are alternative computing paradigms where information is subjected to fluctuations, such that the tools of stochastic thermodynamics are needed to explore their fundamental energy costs \cite{Strasberg-Turing2015,Freitas2021}.

\section{Quantum Energy Initiative}

\subsection{Quantum computing, an overview}
A quantum processor is made of qubits instead of bits, whose states live in Hilbert space and evolution is ruled by the laws of quantum mechanics. The promise of quantum computing is to exploit quantum coherence and entanglement to achieve universal quantum computing with polynomial to exponential gains in complexity with respect to classical computing \cite{Chuang}. Such quantum computational advantage can only show up if decoherence is sufficiently well contained. At the onset of quantum information science in the nineties, the detrimental impact of decoherence triggered fundamental debates of a similar nature as the existence of the Schr\"odinger cat state \cite{QCnightmare}, putting in question the mere existence of quantum computers. Just like for this mythical animal, quantum noise was suspected to destroy massive entanglement, and thus any hope to implement universal, large scale quantum computing.

Fault tolerant quantum error correction schemes drew a path out of this fundamental obstacle \cite{Raussendorf, Devitt, Gottesman}. The quantum threshold theorem states that it is possible to correct errors, even by using noisy gates, provided that the noise level remains below a certain threshold. This allows in principle to execute large scale quantum computations, making it possible to solve large optimization problems, chemistry simulations, or break large RSA keys. The drawback is that it requires huge overhead of physical qubits and classical information processing. However, these theoretical results have motivated an intense experimental work inside academic labs, then in the private sector, in order to design and fabricate qubits of increasing quality and number. 

Owing to this huge experimental effort, small noisy quantum computations have recently been realized. A milestone was reached in 2019 with Google's Sycamore quantum processor made of 53 functional superconducting qubits \cite{Arute}, followed by boson sampling experiments involving a few tens of photons \cite{Pan2020}. Such experiments are dubbed Noisy Intermediate Scale Quantum computations (NISQ) \cite{NISQ}. NISQ processes encompass both gate-based and analog computing, e.g. quantum annealers and quantum simulators. Quantum annealing is a kind of adiabatic quantum computing \cite{AQC}, where the quantum data register is prepared in the ground state of a Hamiltonian, which is externally tuned via the control parameters. Adiabatic evolution of the data register brings it in the state encoding the solution of the computation. Among the envisioned use cases of NISQ platforms, let us mention the resolution of small optimization problems, quantum machine learning and chemistry simulations. Important efforts are currently undertaken to develop useful quantum algorithms on these platforms, and create a possible quantum advantage \cite{Kwek}. 

\subsection{Resource costs of quantum computing}

At the present time, the question of the energetic consumption of quantum computing is barely treated -- the resources usually taken into account for quantum computing being rather the number of physical qubits and cables, or the cryogenic costs -- if there are. In this section I provide a quick overview of the global resources needed to build a quantum computer and how they may translate into an energetic bill. Establishing this bill and minimize it is one of the purposes of the QEI as detailed below.

To render a fair account of the resource consumption of a quantum computer, one should first recall what it is made of. Qubits can be defined by single transitions in natural or artificial atoms, spins, or molecules, polarized photons...Whatever the quantum hardware, the first move is to isolate such quantum levels, and be able to prepare them in well-defined states, coherently manipulate and entangle them, and to finally measure them. This capacity to dig ``poaches of quantum" inside the natural world and address the qubits they contain defines the biggest resource cost. Let us leave aside the cost to fabricate high quality qubits, and to build the physical context around them, and focus on the costs incurred during the computation itself. They encompass the classical drives to manipulate the qubits (laser or micro-wave pulses, magnets, passive and active electronics...) and the physical means to bring them near the quantum processor (wires, interconnections, optical fibers...), the macroscopic means to isolate the processor from the environment (cryostat, shielding, vacuum..), and the qubit readout (amplifiers, superconducting detectors, Charge Coupled Devices...). Below I refer to this ensemble of expenditures as the cost of macroscopic control. 

Let us face it: however hard we try, it is impossible to perfectly isolate the quantum processor, for the simple reason we need to control it and extract information from it. Hence even after having paid the macroscopic control bill, the quantum processor is still described as a noisy quantum system. The computation consists in preparing a set of noisy qubits in a reference state, apply a time-dependent Hamiltonian and perform classical measurements. The computation performance (e.g. fidelity, probability of success...) is lowered by the noise, which can be of thermal origin, but can also involve pure dephasing, be colored, squeezed, non-Markovian... \cite{noise}. A first set of strategies is developed in the field of quantum control, to optimize quantum tasks in the presence of natural noise: in particular, it explores how fast one can drive quantum systems, leading to the derivation of e.g. quantum speed limits \cite{QSL} and shortcuts to adiabaticity \cite{STA}.  These strategies come with fundamental resource costs, below referred to as the cost of quantum control. Another set of strategies directly targets the processor's environment and aims to lower the noise level, by evacuating its entropy. Heat management at the quantum and nano-scale levels is the most natural way (see below). Reservoir engineering \cite{Zoller} gathers an ensemble of techniques modifying the qubits environment and the noise afflicting them. It ultimately leads to engineer the qubits themselves, as it is the case with cat-qubits \cite{Mazyar}. 

Finally, the noise entropy can be lowered by encoding quantum information on larger Hilbert spaces, such that errors can be detected and corrected -- which is the essence of quantum error correction. Depending on the code type (concatenated, surface, colored, bosonic...), fault tolerant quantum processors may involve large numbers of physical qubits to address individually, and important flows of information to process --both resources generating heat that must eventually be evacuated. Note that quantum error correction usually relies on classical information processing, such that the processor description involves a classical and a quantum shell of description. Only autonomous quantum error correction holds the promise to maintain the description at the sole quantum level \cite{Girvin}.

\subsection{Objectives, challenges, methodology}

The shells of description described above are sketched in Fig.\ref{f:QEI}. The management of resources at the full-stack, macroscopic level is ensured by enabling technologies. It relies on effective models that are technology- and hardware dependent. Conversely, fundamental research investigates quantum and nano-scale levels which are largely system-independent. It is at this level that the ultimate assessment between noise and resource cost is made, determining the final computing performance. The management of resources at these levels is ensured by quantum and information thermodynamics, mesoscopic physics and thermoelectricity, quantum algorithmics and computer sciences.

Focusing now on energy, it appears from the description above that the issue of energy management in quantum computing is extremely challenging because of its interdisciplinary nature. The study cannot be restricted to the fundamental level, because it provides no access to the macroscopic energy costs. Reciprocally, the sole macroscopic approach remains blind to the computing performances: in other words, we don't know what we are paying for. As a strong motivation to strongly couple the fundamental and the macroscopic level, let us stress that cost management gives rise to the search for non-trivial sweet spots: increasing the performance requires to lower the noise, lowering the noise requires to increase the control, hence the resources, but increasing the resources generates more noise, which degrades the performance. Optimizations must be jointly conducted by a large range of disciplines: their articulation is the Quantum Energy Initiative called in the present Perspective. For quantum computing, here are a few scientific and technological goals sich an initiative could address:

\begin{itemize}
\item At the fundamental level: derive the fundamental bounds of quantum computing and cooling, explore how energy costs scale with the size of the processor, explore advanced scenarios for energy savings (autonomous error correction, reservoir engineering, reversible computing, quantum energy devices for energy storage and recycling...) 
\item At the macroscopic level: optimize e.g. cryogenic costs, wire conduction, multiplexing, interconnections...
\item Connect the fundamental and the macroscopic levels: how does a target quantum
computing performance impact macroscopic resource consumption? Reciprocally, how do
constraints on macroscopic resources impact quantum computing performances? 
\item Propose guidelines to minimize the global energy consumption in a quantum computation with a given performance; reciprocally, propose guidelines to maximize computing performances under resource constraints.
\item Define energy-based metrics, such as a quantum computing efficiency. Use these metrics to benchmark different qubit technologies (quantum hardware), different computing architectures, codes and compilers (software). 
\item Define the quantum energy advantage and the conditions to draw it out. 
\end{itemize}

\section{State of the art and first results}
In this last section I present a corpus of results already obtained within the various communities, that can serve as seeds to grow the QEI.

\subsection{Fundamental Level} 
Quantum thermodynamics is the natural place to address the fundamental limits of energy consumption at the quantum level. Lying at the crossroad between quantum information science, quantum physics, and stochastic thermodynamics, it aims to extend the concepts of thermodynamics in the quantum realm \cite{Janet-Sai,Binder_book}. One of its primary motivations has been the study of quantum coherence and entanglement as energetic resources. Quantum thermodynamics was formalized as a new kind of resource theory \cite{RMP2019}, searching to exhibit the work value of quantum coherence \cite{korzekwaExtractionWorkQuantum2016,Kammerlander2016,Lostaglio2015,korzekwaExtractionWorkQuantum2016,RMP_Coherence} or quantum advantages in quantum heat engines \cite{Scovil59,Alicki1979,Scully2003,Uzdin2015,Kurizki,Ronnie2016}. Quantum information thermodynamics \cite{Goold_2016} investigates how quantum features impacts the links between information, entropy and energy -- shedding new light on Landauer's bound \cite{Quantum_Landauer}, the thermodynamic cost of quantum operations \cite{Owen}, and quantum Maxwell demons \cite{Huard, Murch, Masuyama18}. Quantum stochastic thermodynamics \cite{Strasberg_Book} brings these investigations down to the level of stochastic quantum trajectories. Fluctuation theorems are extended in the quantum regime (See \cite{Hanggi} and the Chapter ``Quantum fluctuation theorems" in \cite{Binder_book}), possibly involving measurement and feedback processes \cite{Funo,Masuyama18}. Thermodynamic Uncertainty Relations (TUR) \cite{TUR} that constraint power, fluctuations, and entropic cost are also expected to impact the cost of quantum tasks. The derivation of fundamental quantum bounds requires to understand and quantify irreversibility in the quantum world \cite{Paternostro}.

Directly related to fundamental energy costs of quantum technologies, an important corpus of results has been obtained on the energetics of quantum computing \cite{Gea}, of single qubit gates \cite{Mottonen,Barbieri,Ben21,Deffner}, of quantum measurements \cite{Huber},  the relation between noise and performances in quantum amplifiers \cite{Caves82} and communication channels \cite{Caves94}, thermodynamic analyzes of error correcting codes as Maxwell demons \cite{MD_QEC, Landi} and adiabatic quantum computing \cite{Campisi,Campisi21}. Quantum batteries are promising devices to store and retrieve energy on demand at the quantum level, with a predicted impact of quantum coherence and correlations on charging powers and efficiencies \cite{Campaioli2017,Ferraro2018, Quantacell,Polini2019,Julia-Farre2020,Monsel2020,Giovannetti2021,Caravelli2021}. Heat management at the nano and mesoscopic level is the field of expertise of thermo-electricity, that optimizes the performances of quantum energy devices such as thermal diodes and refrigerators \cite{Pekola_RMP_2006,Jordan_2014,Pekola_2015,Whitney2017} towards fundamental limits of cooling \cite{Freitas2017}. 

Importantly in the last ten years, the community enrolled an increasing number of experimenters, leading to the definition of operational concepts associated to measurable quantities. Quantum Maxwell's demons, quantum engines, fine thermometry experiments, measurement of entropy production were implemented on a wide range of experimental platforms: ion traps \cite{Poschinger,Matsukevich,Singer}, single electron transistors \cite{Pekola}, quantum photonics \cite{Barbieri}, Nitrogen-Vacancy centers \cite{Walmsley,Campisi21b}, Rydberg atoms \cite{Dotsenko}, superconducting circuits \cite{Huard, Murch, Gasparinetti, Fedorov, Masuyama18}, Nuclear Magnetic Resonance \cite{Serra}, cold atoms \cite{Schmiedmayer, Donner}, levitated nanoparticles \cite{Kiesel}, SiN membranes \cite{Ares}, including working quantum processors \cite{Campisi21c}.

\subsection{Relating fundamental to full-stack}
As presented above, the QEI relies on the creation of connections between the fundamental and the full-stack level. We have created some of these connections between quantum computing, quantum thermodynamics and cryo-electronics, giving rise to the results reported in \cite{Marco_2021, Marco_2022, Marco_PhD}. Our first task was to set up a common language, focusing on the most important concepts and their relations: noise, resource and metric of performance. The methodology is general: noise can be any kind of quantum noise, and the resource any available physical resource (cryogenic power, number of qubits, fidelity of the qubits, multiplexing, complexity of decoding errors...). Metrics of performance for quantum computing can be e.g. the fidelity (average or minimal), or the probability of success of the algorithm. As explained above, noise and metric of performance are characteristics of the fundamental level. Conversely each level of description is associated to a specific resource cost, e.g. energy or power -- predominant costs being billed at the macroscopic level. At this level of generality and if they are properly listed and quantified, noise, resources and metric are expected to abide by the following principles:

\begin{enumerate}
\item {\it Unbounded resources principle: }for a fixed noise rate, arbitrary large metrics of success can be reached by consuming arbitrary resources in arbitrary large amounts. 
\item {\it Resource minimization principle: }for a fixed noise rate and target metric of success, there is a minimum resource cost.
\item {\it Resource limitation principle: }for a fixed noise rate and resource cost, there is a maximal metric of success.
 \end{enumerate}

We have dubbed this framework MNR as for Metric, Noise, Resource. At the fundamental level, when the noise is thermal, and only energetic resources are considered, MNR exactly matches the thermodynamical framework, ``resources" corresponding to ``work" and the resource minimization principle to the existence of a fundamental bound. In a full-stack approach, MNR provides effective and intuitive conceptual tools to optimize global energy costs. MNR thus serves as a proper soil for crossed fertilization. Each MNR principle stated above defines a joint roadmap for both fundamental research and technological developments. The unbounded resources principle captures the mindset that has prevailed so far, whose motto ``Maximize the performance, whatever the cost" is less and less sustainable. Conversely, the resource minimization principle and the resource limitation principle quantitatively take into account the impact of the resource cost. They lay the ground for a desirable paradigm shift promoted by the present Perspective.

\subsubsection{Resource limitation}

We have explored to which extent resource constraints impact the quantum threshold theorem \cite{Marco_2021}. As a matter of fact, this cornerstone of fault tolerant quantum computing was derived without resource limitation, apart from the one set by the qubit natural fidelity. It states that arbitrary accuracies can always be reached, at the price of always larger overheads of physical qubits to encode a logical qubit -- provided that the noise afflicting the quantum gates remains below a certain threshold. But putting a resource constraint at the full-stack level (such as the available cooling budget, or the size of the quantum processor) generates an effective scale-dependent noise that impacts the processes at the fundamental level. In particular, we have evidenced that upper bounding the available resources leads to a breakdown of the mere concept of threshold: there is an optimal ``code size", i.e. number of physical qubits per logical qubit, that maximizes the metric of performance -- and beyond which more error correction degrades the computation accuracy. This gives a glimpse of the massive change of landscape to expect as soon as resources limitations will start being taken into account -- and the extremely stimulating questions that will emerge from this new situation.

\subsubsection{Resource cost minimization}

Conversely, we have studied the energetic behavior of a full-stack, fault-tolerant quantum computer made of superconducting qubits \cite{Marco_PhD,Marco_2022}. The model involves the classical control (control electronics, amplifiers, attenuators), the wires conduction, different cryogenic stages, and realistic qubits models. The resource to optimize is the power consumption at the macroscopic level, which encompasses the cryogeny and the classical control. In a nutshell, the connection between the microscopic and the macroscopic standpoint is captured by the simple relation:

\begin{equation}
P=\frac{T}{T_q}\dot{Q},
\end{equation} 

where $T$ (resp. $T_q$) stands for the room temperature (resp. the quantum processor temperature) and $P$ is the power consumed by the cryogeny. The large ratio $T/T_q$ acts as a magnification factor between the heat generation at the quantum level $\dot{Q}$ and the cryo-power consumption $P$ at the full-stack level. Two mechanisms are competing: less noise mandates a bigger code and more physical qubits, but more physical qubits give rise to more heat generation $\dot{Q}$, hence more noise -- motivating the search for an optimized power consumption. The study reveals the strong impact of the classical control electronics and of the qubit quality on the global energetic performance. In particular, our extensive account of resources provides us with realistic estimations for the cryo-power consumption to break a 2048 bits RSA key, which could get reduced to a few tens of kW after optimizing the macroscopic cost \cite{Marco_2022} and use an optimized surface code \cite{Gidney}. Actually in this situation, the energy cost appears to be largely dominated by the classical information processing, necessary to decode errors -- another argument for a global, interdisciplinary treatment of energetic issues.

\subsubsection{Quantum computing efficiency}
MNR brings out quantitative and possibly analytic relations between computing performance and resource consumption, at all levels between the fundamental and the full-stack. This ability invites to define a quantum computing (energetic) efficiency $\eta = {\cal M}/{\cal R}$, where ${\cal R}$ stands for the (energetic) resource cost and ${\cal M}$ for the metric of success. $\eta$ plays a quite different role from figures of merit previously proposed for quantum computing, such as the quantum volume or the Q-score. These metrics capture the capacity of a given qubit technology to perform complex algorithms with a good accuracy -- whatever the cost. Conversely, $\eta$ captures the sustainability of a computing strategy, and can be seen as a quantum equivalent of the "Performance per Watt" used in the classical computing paradigm.  Just like the most energy efficient supercomputers give rise to the ``Green500" ranking, the quantum computing efficiency carries the seeds for an upcoming QuGreen 500 ranking.

Like its classical counterpart, the quantum computing efficiency is versatile and allows to benchmark both the hardware and the software. On the one hand, it provides a fair way to compare different technologies of qubits, whether at the level of bare quantum gates and circuits, or in a ``dressed", full-stack picture. In particular, it will be interesting to study how photonic and ion-based quantum computing compares to solid-state quantum computing, and if the less stringent cryogenic needs make photons and ions natural candidates for energy efficient quantum computing.  On the other hand for a given hardware and a given quantum algorithm, $\eta$ allows to benchmark different computing architectures, e.g. various circuits, codes size and connectivity. $\eta$ is a component of the score card assessing the potential for scalability.

\section{Discussion}

\subsubsection{Quantum energy advantage}
Let us first recall that the primary meaning of the quantum advantage is computational: it is said to be reached if a quantum processor performs a task faster than its hypothetical classical counterpart, usually a supercomputer programmed to solve the same problem with the best in-class algorithm for each architecture. The quantum energy advantage is defined by drawing the energetic consequences of the computational one: if a quantum processor requires fewer physical operations than its classical counterpart to implement a given algorithm, then an equivalent advantage should appear in its energetic consumption. This guess was phenomenologically confirmed by comparing the power consumed by Google's Sycamore processor and IBM Summit supercomputer \cite{Arute}. The present Perspective demonstrates that minimal energy costs at the full-stack level do not simply scale like the number of operations, and require non-trivial optimizations -- only after which a clear quantum advantage can be claimed.

From a fundamental point of view, justifying the existence of a quantum advantage and quantifying it is a tricky and fascinating problem. Firstly, the ``classical counterpart" is a moving target since classical algorithms and systems architectures can benefit from various optimizations. This is an ongoing debate \cite{Waintal}. Secondly on the quantum side, the choice of qubits technology, computing architecture, classical control, impacts the energetic performance. In the same way, the choice of technology (especially its potential for reversibility, whether thermodynamical or logical) impacts the minimal energy cost of classical computing. As a consequence, it would be fair to compare technology-independent energy costs, i.e. classical and quantum bounds -- whose derivation is still to come. Demonstrating a fundamental quantum energy advantage thus brings us back to hitting fundamental bounds. It is the final destination of the roadmap for resource minimization, where all intermediate steps will generate huge progresses in terms of resource management and sustainability.

\subsubsection{Societal and technological impact }

The emergence of quantum technologies is a fantastic opportunity to create new circles of innovation between fundamental research and technological developments. The QEI is one of these circles or ``short loops" . The present Perspective demonstrates how such a QEI is a necessity to bring out energy efficient quantum technologies, by optimizing their resources consumption along their development. While I solely treated the case of quantum computing, the missions of the QEI are transversal, impacting all branches of quantum technologies (quantum communication, quantum simulation, quantum sensing) as well as classical informations technologies.

I have voluntarily left aside the potential of quantum computers to solve energetic problems and bring solutions to the climate change \cite{Qforclimate}. These considerations, however, must be taken into account by society to assess how much it wants to invest in quantum technologies. In the same way, I have not elaborated on the ``rebound effect", i.e. the well-identified mechanism by which gains in the energetic efficiency of a technology tend to translate into more needs, increasing the global energetic consumption. This is a societal and political issue which is beyond the scope of this Perspective. As scientists and technologists, our work is to provide physically sound options to allow society makes enlightened choices, and that's what a quantum energy initiative could deliver.

My final words are to emphasize the timeliness of a QEI. I hear too often in discussions that one should first build a useful quantum computer, or the quantum internet, and show that these technologies work before caring about their energetic consumption. It won't come as a surprise, but I don't subscribe to this view. Optimizing resources to reach well-defined performances guarantees the possibility to make smart technological choices while the technology is being built and before dead-ends show up. It also allows end users to know what performance they are exactly paying for, and to know how to get an optimal price. On the contrary, being blind to energetic consumption is the surest way to get stuck in dead ends and waste money -- especially on the road to universal quantum computing where unoptimized energetic costs may quickly reach GigaWatt levels \cite{Marco_2022}. 

Our recent works show that looking at the energetics of quantum computing from a multi-disciplinary perspective reveals new design principles, optimization and benchmarking techniques which bring new and essential milestones on the road to scalability. Our collaboration involves experts in fault-tolerant quantum computing, quantum thermodynamics, solid-state physics, cryo-CMOS. It already attracted experimenters from the superconducting, silicon and photon qubits platforms. The initiative should now grow to involve the full range of quantum hardware, and give rise to experiments where energy-based metrics are actually optimized and measured. It should quickly enroll software researchers and developers, who will use energetic considerations to benchmark and optimize quantum computing architectures, codes and compilers. As mentioned in the introduction, it should also extend its scope to all pillars of quantum technologies, i.e. metrology and communication. Ultimately, energetic workpackages should appear in research and development projects, energy departments in startups and companies. 

Behind these considerations, there is a much deeper societal reason. Quantum technologies are very young, and are just entering the economical sphere. As ``deep techs", they are still strongly coupled to fundamental research and rely a lot on public funding, such that there is flexibility on priority adjustments. Integrating the energetic dimension now is possible and will send a strong signal to society. It would ensure that science and technology stakeholders care about bringing out responsible innovations \cite{hype}, by developing conceptual tools and applying them to minimize the resource consumption. It will make quantum technologies a model of virtuous deployment process for future innovations. This is essential in a world where resources are limited, and where the positive impact of science and technology for mankind must be demonstrated more than ever.

\section{Acknowledgments}
The author warmly thanks P. Camati, M. Campisi, S. Ciliberto, S. Deffner, M. Devoret, M. Esposito, M. Fellous-Asiani, M. Frank, P. Grangier, D. Herrera-Mart\'i, P. Hilaire, J. Parrondo, M. Pollini, H. Ribot, P. Strasberg, Y. Thonnart, M. Ueda, M. Vinet and R. Whitney for their fruitful comments, and O. Ezratty for his continuous feedback and support. The author's research is funded by the Foundational Questions Institute Fund (Grant No.FQXi-IAF19-01 and FQXi-IAF19-05), the Templeton World Charity Foundation, Inc. (Grant No. TWCF0338), the ANR Research Collaborative Project ``Qu-DICE" (ANR-18-CE47-0009) and the European Union Horizon 2020 research and innovation programme under the collaborative project QLSI (Grant agreement No 951852).


\begin{thebibliography}{117}%
\makeatletter
\providecommand \@ifxundefined [1]{%
 \@ifx{#1\undefined}
}%
\providecommand \@ifnum [1]{%
 \ifnum #1\expandafter \@firstoftwo
 \else \expandafter \@secondoftwo
 \fi
}%
\providecommand \@ifx [1]{%
 \ifx #1\expandafter \@firstoftwo
 \else \expandafter \@secondoftwo
 \fi
}%
\providecommand \natexlab [1]{#1}%
\providecommand \enquote  [1]{``#1''}%
\providecommand \bibnamefont  [1]{#1}%
\providecommand \bibfnamefont [1]{#1}%
\providecommand \citenamefont [1]{#1}%
\providecommand \href@noop [0]{\@secondoftwo}%
\providecommand \href [0]{\begingroup \@sanitize@url \@href}%
\providecommand \@href[1]{\@@startlink{#1}\@@href}%
\providecommand \@@href[1]{\endgroup#1\@@endlink}%
\providecommand \@sanitize@url [0]{\catcode `\\12\catcode `\$12\catcode
  `\&12\catcode `\#12\catcode `\^12\catcode `\_12\catcode `\%12\relax}%
\providecommand \@@startlink[1]{}%
\providecommand \@@endlink[0]{}%
\providecommand \url  [0]{\begingroup\@sanitize@url \@url }%
\providecommand \@url [1]{\endgroup\@href {#1}{\urlprefix }}%
\providecommand \urlprefix  [0]{URL }%
\providecommand \Eprint [0]{\href }%
\providecommand \doibase [0]{https://doi.org/}%
\providecommand \selectlanguage [0]{\@gobble}%
\providecommand \bibinfo  [0]{\@secondoftwo}%
\providecommand \bibfield  [0]{\@secondoftwo}%
\providecommand \translation [1]{[#1]}%
\providecommand \BibitemOpen [0]{}%
\providecommand \bibitemStop [0]{}%
\providecommand \bibitemNoStop [0]{.\EOS\space}%
\providecommand \EOS [0]{\spacefactor3000\relax}%
\providecommand \BibitemShut  [1]{\csname bibitem#1\endcsname}%
\let\auto@bib@innerbib\@empty
%</preamble>

\bibitem [{\citenamefont {Deutsch}(2020)}]{Deutsch_PRXQuantum}%
  \BibitemOpen
  \bibfield  {author} {\bibinfo {author} {\bibfnamefont {I.~H.}\ \bibnamefont
  {Deutsch}},\ }\bibfield  {title} {\bibinfo {title} {Harnessing the power of
  the second quantum revolution},\ }\href
  {https://doi.org/10.1103/PRXQuantum.1.020101} {\bibfield  {journal} {\bibinfo
   {journal} {PRX Quantum}\ }\textbf {\bibinfo {volume} {1}},\ \bibinfo {pages}
  {020101} (\bibinfo {year} {2020})}\BibitemShut {NoStop}%
  \bibitem [{\citenamefont {Puebla}\ \emph {et~al.}(2020)\citenamefont {Puebla},
  \citenamefont {Kim}, \citenamefont {Kondou},\ and\ \citenamefont
  {Otani}}]{Puebla}%
  \BibitemOpen
  \bibfield  {author} {\bibinfo {author} {\bibfnamefont {J.}~\bibnamefont
  {Puebla}}, \bibinfo {author} {\bibfnamefont {J.}~\bibnamefont {Kim}},
  \bibinfo {author} {\bibfnamefont {K.}~\bibnamefont {Kondou}},\ and\ \bibinfo
  {author} {\bibfnamefont {Y.}~\bibnamefont {Otani}},\ }\bibfield  {title}
  {\bibinfo {title} {Spintronic devices for energy-efficient data storage and
  energy harvesting},\ }\bibfield  {journal} {\bibinfo  {journal} {{
  Communications Materials}}\ }\textbf {\bibinfo {volume} {1}},\ \href
  {https://doi.org/10.1038/s43246-020-0022-5} {10.1038/s43246-020-0022-5}
  (\bibinfo {year} {2020})\BibitemShut {NoStop}%
  \bibitem [{\citenamefont {Gammaitoni}\ \emph {et~al.}(2015)\citenamefont
  {Gammaitoni}, \citenamefont {Chiuchi{\'{u}}}, \citenamefont {Madami},\ and\
  \citenamefont {Carlotti}}]{Gammaitoni_2015}%
  \BibitemOpen
  \bibfield  {author} {\bibinfo {author} {\bibfnamefont {L.}~\bibnamefont
  {Gammaitoni}}, \bibinfo {author} {\bibfnamefont {D.}~\bibnamefont
  {Chiuchi{\'{u}}}}, \bibinfo {author} {\bibfnamefont {M.}~\bibnamefont
  {Madami}},\ and\ \bibinfo {author} {\bibfnamefont {G.}~\bibnamefont
  {Carlotti}},\ }\bibfield  {title} {\bibinfo {title} {Towards zero-power
  {ICT}},\ }\href {https://doi.org/10.1088/0957-4484/26/22/222001} {\bibfield
  {journal} {\bibinfo  {journal} {Nanotechnology}\ }\textbf {\bibinfo {volume}
  {26}},\ \bibinfo {pages} {222001} (\bibinfo {year} {2015})}\BibitemShut
  {NoStop}%
  \bibitem [{\citenamefont {Arute}\ \emph {et~al.}(2019)\citenamefont {Arute},
   \citenamefont {Neven},\ and\ \citenamefont {Martinis}}]{Arute}%
  \BibitemOpen
  \bibfield  {author} {\bibinfo {author} {\bibfnamefont {F.}~\bibnamefont
  {Arute}},  \bibinfo {author} {\bibfnamefont {et.}\
  \bibnamefont {al}},\ }
  \bibfield  {title} {\bibinfo {title} {Quantum
  supremacy using a programmable superconducting processor},\ }\href
  {https://doi.org/10.1038/s41586-019-1666-5} 
  {\bibfield  {journal} {\bibinfo {journal} {Nature}\ }\textbf {\bibinfo {volume} {574}},\ \bibinfo {pages}
  {505} (\bibinfo {year} {2019})}
  \BibitemShut {NoStop}%
  \bibitem [{\citenamefont {Curzon}\ and\ \citenamefont
  {Ahlborn}(1975)}]{CA_1975}%
  \BibitemOpen
  \bibfield  {author} {\bibinfo {author} {\bibfnamefont {F.~L.}\ \bibnamefont
  {Curzon}}\ and\ \bibinfo {author} {\bibfnamefont {B.}~\bibnamefont
  {Ahlborn}},\ }\bibfield  {title} {\bibinfo {title} {Efficiency of a carnot
  engine at maximum power output},\ }\bibfield  {journal} {\bibinfo  {journal}
  {American Journal of Physics}\ }\textbf {\bibinfo {volume} {43}},\ \href
  {https://doi.org/10.1119/1.10023} {10.1119/1.10023} (\bibinfo {year}
  {1975})\BibitemShut {NoStop}%
\bibitem [{\citenamefont {Seifert}(2008)}]{Seifert2008}%
  \BibitemOpen
  \bibfield  {author} {\bibinfo {author} {\bibfnamefont {U.}~\bibnamefont
  {Seifert}},\ }\bibfield  {title} {\bibinfo {title} {Stochastic
  thermodynamics: Principles and perspectives.},\ }\href@noop {} {\bibfield
  {journal} {\bibinfo  {journal} {{ Eur. Phys. J. B }}\ }\textbf {\bibinfo
  {volume} {64}},\ \bibinfo {pages} {423} (\bibinfo {year} {2008})}\BibitemShut
  {NoStop}%
\bibitem [{\citenamefont {Peliti}\ and\ \citenamefont
  {Pigolotti}(2021)}]{Peliti}%
  \BibitemOpen
  \bibinfo {editor} {\bibfnamefont {L.}~\bibnamefont {Peliti}}\ and\ \bibinfo
  {editor} {\bibfnamefont {S.}~\bibnamefont {Pigolotti}},\ eds.,\ \href@noop {}
  {\emph {\bibinfo {title} {Stochastic Thermodynamics: An Introduction}}}\
  (\bibinfo  {publisher} {{Princeton}},\ \bibinfo {year} {2021})\BibitemShut
  {NoStop}%
\bibitem [{\citenamefont {Sekimoto}(2010)}]{Sekimoto}%
  \BibitemOpen
  \bibinfo {editor} {\bibfnamefont {K.}~\bibnamefont {Sekimoto}},\ ed.,\
  \href@noop {} {\emph {\bibinfo {title} {Stochastic Energetics}}},\ Lecture
  Notes in Physics\ (\bibinfo  {publisher} {{Springer}},\ \bibinfo {year}
  {2010})\BibitemShut {NoStop}%
\bibitem [{\citenamefont {Klages}\ \emph {et~al.}()\citenamefont {Klages},
  \citenamefont {Just}, \citenamefont {Jarzynski},\ and\ \citenamefont
  {Schuster}}]{CJ_Book}%
  \BibitemOpen
  \bibinfo {editor} {\bibfnamefont {R.}~\bibnamefont {Klages}}, \bibinfo
  {editor} {\bibfnamefont {W.}~\bibnamefont {Just}}, \bibinfo {editor}
  {\bibfnamefont {C.}~\bibnamefont {Jarzynski}},\ and\ \bibinfo {editor}
  {\bibfnamefont {H.}~\bibnamefont {Schuster}},\ eds.,\ \href@noop {} {\emph
  {\bibinfo {title} {Nonequilibrium Statistical Physics of Small Systems:\\}
  {Fluctuation Relations and Beyond}}}\ (\bibinfo  {publisher}
  {{Wiley}})\BibitemShut {NoStop}%
\bibitem [{\citenamefont {Hill}()}]{Hill}%
  \BibitemOpen
  \bibinfo {editor} {\bibfnamefont {T.-L.}\ \bibnamefont {Hill}},\ ed.,\
  \href@noop {} {\emph {\bibinfo {title} {Thermodynamics of small systems}}}\
  (\bibinfo  {publisher} {{Dover}})\BibitemShut {NoStop}%
\bibitem [{\citenamefont {Bennett}(1982)}]{Bennett82}%
  \BibitemOpen
  \bibfield  {author} {\bibinfo {author} {\bibfnamefont {C.~H.}\ \bibnamefont
  {Bennett}},\ }\bibfield  {title} {\bibinfo {title} {The thermodynamics of
  computation, A review},\ }\href@noop {} {\bibfield  {journal} {\bibinfo
  {journal} {{ Int. J. Theor. Phys. }}\ }\textbf {\bibinfo {volume} {21}},\
  \bibinfo {pages} {905940} (\bibinfo {year} {1982})}\BibitemShut {NoStop}%
\bibitem [{\citenamefont {Landauer}(1961)}]{Landauer61}%
  \BibitemOpen
  \bibfield  {author} {\bibinfo {author} {\bibfnamefont {R.}~\bibnamefont
  {Landauer}},\ }\bibfield  {title} {\bibinfo {title} {Irreversibility and heat
  generation in the computing process},\ }\href
  {https://doi.org/10.1147/rd.53.0183} {\bibfield  {journal} {\bibinfo
  {journal} {{ IBM Journal of Research and Development }}\ ,\ \bibinfo {pages}
  {183}} (\bibinfo {year} {1961})}\BibitemShut {NoStop}%
\bibitem [{\citenamefont {Sagawa}(2014)}]{Sagawa14}%
  \BibitemOpen
  \bibfield  {author} {\bibinfo {author} {\bibfnamefont {T.}~\bibnamefont
  {Sagawa}},\ }\bibfield  {title} {\bibinfo {title} {Thermodynamic and logical
  reversibilities revisited},\ }\href
  {https://doi.org/10.1088/1742-5468/2014/03/p03025} {\bibfield  {journal}
  {\bibinfo  {journal} {Phys. Rev. Lett.}\ }\textbf {\bibinfo {volume}
  {2014}},\ \bibinfo {pages} {P03025} (\bibinfo {year} {2014})}\BibitemShut
  {NoStop}%
\bibitem [{\citenamefont {Sagawa}\ and\ \citenamefont
  {Ueda}(2009)}]{Sagawa-Ueda09}%
  \BibitemOpen
  \bibfield  {author} {\bibinfo {author} {\bibfnamefont {T.}~\bibnamefont
  {Sagawa}}\ and\ \bibinfo {author} {\bibfnamefont {M.}~\bibnamefont {Ueda}},\
  }\bibfield  {title} {\bibinfo {title} {Minimal energy cost for thermodynamic
  information processing: Measurement and information erasure},\ }\href
  {https://doi.org/10.1103/PhysRevLett.102.250602} {\bibfield  {journal}
  {\bibinfo  {journal} {Phys. Rev. Lett.}\ }\textbf {\bibinfo {volume} {102}},\
  \bibinfo {pages} {250602} (\bibinfo {year} {2009})}\BibitemShut {NoStop}%
\bibitem [{\citenamefont {Maruyama}\ \emph {et~al.}(2009)\citenamefont
  {Maruyama}, \citenamefont {Nori},\ and\ \citenamefont {Vedral}}]{MD}%
  \BibitemOpen
  \bibfield  {author} {\bibinfo {author} {\bibfnamefont {K.}~\bibnamefont
  {Maruyama}}, \bibinfo {author} {\bibfnamefont {F.}~\bibnamefont {Nori}},\
  and\ \bibinfo {author} {\bibfnamefont {V.}~\bibnamefont {Vedral}},\
  }\bibfield  {title} {\bibinfo {title} {Colloquium: The physics of Maxwell's
  demon and information},\ }\href {https://doi.org/10.1103/RevModPhys.81.1}
  {\bibfield  {journal} {\bibinfo  {journal} {Rev. Mod. Phys.}\ }\textbf
  {\bibinfo {volume} {81}},\ \bibinfo {pages} {1} (\bibinfo {year}
  {2009})}\BibitemShut {NoStop}%
\bibitem [{\citenamefont {Leff}\ and\ \citenamefont {Rex}(2002)}]{Leff-Rex}%
  \BibitemOpen
  \bibinfo {editor} {\bibfnamefont {H.}~\bibnamefont {Leff}}\ and\ \bibinfo
  {editor} {\bibfnamefont {F.}~\bibnamefont {Rex}},\ eds.,\ \href@noop {}
  {\emph {\bibinfo {title} {Maxwell's Demon 2 Entropy, }\\ {Classical and Quantum
  Information, Computing}}}\ (\bibinfo  {publisher} {{CRC Press}},\ \bibinfo
  {year} {2002})\BibitemShut {NoStop}%
\bibitem [{\citenamefont {Theis}\ and\ \citenamefont {Wong}(2017)}]{EndMoore}%
  \BibitemOpen
  \bibfield  {author} {\bibinfo {author} {\bibfnamefont {T.~N.}\ \bibnamefont
  {Theis}}\ and\ \bibinfo {author} {\bibfnamefont {H.-P.}\ \bibnamefont
  {Wong}},\ }\bibfield  {title} {\bibinfo {title} {The end of Moore's law: A
  new beginning for information technology},\ }\href
  {https://doi.org/10.1109/MCSE.2017.29} {\bibfield  {journal} {\bibinfo
  {journal} {Computing in Science and Engineering}\ }\textbf {\bibinfo {volume}
  {19}},\ \bibinfo {pages} {41} (\bibinfo {year} {2017})}\BibitemShut {NoStop}%
\bibitem [{\citenamefont {Ezratty}(2021)}]{Oez}%
  \BibitemOpen
  \bibinfo {editor} {\bibfnamefont {O.}~\bibnamefont {Ezratty}},\ ed.,\
  \href@noop {} {\emph {\bibinfo {title} {Understanding Quantum
  Technologies}}}\ (\bibinfo  {publisher} {{Le Lab Quantique}},\ \bibinfo
  {year} {2021})\BibitemShut {NoStop}%
\bibitem [{\citenamefont {Desislavov}\ \emph {et~al.}(2021)\citenamefont
  {Desislavov}, \citenamefont {Martinez-Plumed},\ and\ \citenamefont
  {Hernandez-Orallo}}]{Desislavov}%
  \BibitemOpen
  \bibfield  {author} {\bibinfo {author} {\bibfnamefont {R.}~\bibnamefont
  {Desislavov}}, \bibinfo {author} {\bibfnamefont {F.}~\bibnamefont
  {Martinez-Plumed}},\ and\ \bibinfo {author} {\bibfnamefont {J.}~\bibnamefont
  {Hernandez-Orallo}},\ }\bibfield  {title} {\bibinfo {title} {Compute and
  energy consumption trends in deep learning inference},\ }\href@noop {}
  {\bibfield  {journal} {\bibinfo  {journal} {arXiv:2109.05472}\ } (\bibinfo {year}
  {2021})}\BibitemShut {NoStop}%
\bibitem [{\citenamefont {Bérut}\ \emph {et~al.}(2012)\citenamefont {Bérut},
  \citenamefont {Arakelyan}, \citenamefont {Petrosyan}, \citenamefont
  {Ciliberto}, \citenamefont {Dillenschneider},\ and\ \citenamefont
  {Lutz}}]{Sergio}%
  \BibitemOpen
  \bibfield  {author} {\bibinfo {author} {\bibfnamefont {A.}~\bibnamefont
  {Berut}}, \bibinfo {author} {\bibfnamefont {A.}~\bibnamefont {Arakelyan}},
  \bibinfo {author} {\bibfnamefont {A.}~\bibnamefont {Petrosyan}}, \bibinfo
  {author} {\bibfnamefont {S.}~\bibnamefont {Ciliberto}}, \bibinfo {author}
  {\bibfnamefont {R.}~\bibnamefont {Dillenschneider}},\ and\ \bibinfo {author}
  {\bibfnamefont {E.}~\bibnamefont {Lutz}},\ }\bibfield  {title} {\bibinfo
  {title} {Experimental verification of Landauer's principle linking
  information and thermodynamics},\ }\href
  {https://doi.org/10.1038/nature10872} {\bibfield  {journal} {\bibinfo
  {journal} {{Nature}}\ }\textbf {\bibinfo {volume} {483}},\ \bibinfo {pages}
  {189} (\bibinfo {year} {2012})}\BibitemShut {NoStop}%
\bibitem [{\citenamefont {Anantharam}\ \emph {et~al.}(2004)\citenamefont
  {Anantharam}, \citenamefont {He}, \citenamefont {Natarajan}, \citenamefont
  {Xie},\ and\ \citenamefont {Frank}}]{ACMOS}%
  \BibitemOpen
  \bibfield  {author} {\bibinfo {author} {\bibfnamefont {V.}~\bibnamefont
  {Anantharam}}, \bibinfo {author} {\bibfnamefont {M.}~\bibnamefont {He}},
  \bibinfo {author} {\bibfnamefont {K.}~\bibnamefont {Natarajan}}, \bibinfo
  {author} {\bibfnamefont {H.}~\bibnamefont {Xie}},\ and\ \bibinfo {author}
  {\bibfnamefont {M.}~\bibnamefont {Frank}},\ }\bibfield  {title} {\bibinfo
  {title} {Driving fully-adiabatic logic circuits using custom high-Q MEMS
  resonators.}\ }(\bibinfo {year} {2004})\ pp.\ \bibinfo {pages}
  {5--11}\BibitemShut {NoStop}%
\bibitem [{\citenamefont {Bennett}(1973)}]{Bennett73}%
  \BibitemOpen
  \bibfield  {author} {\bibinfo {author} {\bibfnamefont {C.~H.}\ \bibnamefont
  {Bennett}},\ }\bibfield  {title} {\bibinfo {title} {Logical reversibility of
  computation},\ }\href@noop {} {\bibfield  {journal} {\bibinfo  {journal} {{
  IBM J. Res. Dev.}}\ }\textbf {\bibinfo {volume} {17}},\ \bibinfo {pages}
  {525?532} (\bibinfo {year} {1973})}\BibitemShut {NoStop}%
\bibitem [{\citenamefont {Fredkin}\ and\ \citenamefont
  {Toffoli}(1982)}]{Fredkin_Toffoli}%
  \BibitemOpen
  \bibfield  {author} {\bibinfo {author} {\bibfnamefont {E.}~\bibnamefont
  {Fredkin}}\ and\ \bibinfo {author} {\bibfnamefont {T.}~\bibnamefont
  {Toffoli}},\ }\bibfield  {title} {\bibinfo {title} {Conservative logic.},\
  }\href {https://doi.org/10.1007/BF01857727} {\bibfield  {journal} {\bibinfo
  {journal} {{ International Journal of Theoretical Physics}}\ }\textbf
  {\bibinfo {volume} {21}},\ \bibinfo {pages} {219} (\bibinfo {year}
  {1982})}\BibitemShut {NoStop}%
\bibitem [{\citenamefont {Strasberg}\ \emph {et~al.}(2015)\citenamefont
  {Strasberg}, \citenamefont {Cerrillo}, \citenamefont {Schaller},\ and\
  \citenamefont {Brandes}}]{Strasberg-Turing2015}%
  \BibitemOpen
  \bibfield  {author} {\bibinfo {author} {\bibfnamefont {P.}~\bibnamefont
  {Strasberg}}, \bibinfo {author} {\bibfnamefont {J.}~\bibnamefont {Cerrillo}},
  \bibinfo {author} {\bibfnamefont {G.}~\bibnamefont {Schaller}},\ and\
  \bibinfo {author} {\bibfnamefont {T.}~\bibnamefont {Brandes}},\ }\bibfield
  {title} {\bibinfo {title} {Thermodynamics of stochastic turing machines},\
  }\href {https://doi.org/10.1103/PhysRevE.92.042104} {\bibfield  {journal}
  {\bibinfo  {journal} {Phys. Rev. E}\ }\textbf {\bibinfo {volume} {92}},\
  \bibinfo {pages} {042104} (\bibinfo {year} {2015})}\BibitemShut {NoStop}%
\bibitem [{\citenamefont {Frank}(2018)}]{Frank18}%
  \BibitemOpen
  \bibfield  {author} {\bibinfo {author} {\bibfnamefont {M.~P.}\ \bibnamefont
  {Frank}},\ }\bibfield  {title} {\bibinfo {title} {Physical foundations of
  Landauer's principle},\ }in\ \href@noop {} {\emph {\bibinfo {booktitle}
  {Reversible Computation}}},\ \bibinfo {editor} {edited by\ \bibinfo {editor}
  {\bibfnamefont {J.}~\bibnamefont {Kari}}\ and\ \bibinfo {editor}
  {\bibfnamefont {I.}~\bibnamefont {Ulidowski}}}\ (\bibinfo  {publisher}
  {Springer International Publishing},\ \bibinfo {address} {Cham},\ \bibinfo
  {year} {2018})\ pp.\ \bibinfo {pages} {3--33}\BibitemShut {NoStop}%
\bibitem [{\citenamefont {Frank}(2020)}]{Frank20}%
  \BibitemOpen
  \bibfield  {author} {\bibinfo {author} {\bibfnamefont {M.}~\bibnamefont
  {Frank}},\ }\bibfield  {title} {\bibinfo {title} {Architectural, algorithmic,
  and systems engineering issues for reversible computing},\ }in\ \href
  {https://cra.org/ccc/wp-content/uploads/sites/2/2020/10/CCC-Frank-Day3-v3.pdf}
  {\emph {\bibinfo {booktitle} {CCC Workshop on Physics and Engineering Issues}\\
 { in Adiabatic/Reversible Classical Computing}}}\ (\bibinfo {year}
  {2020})\BibitemShut {NoStop}%
\bibitem [{\citenamefont {Takeuchi}\ \emph {et~al.}(2017)\citenamefont
  {Takeuchi}, \citenamefont {Yamanashi},\ and\ \citenamefont
  {Yoshikawa}}]{RQFP}%
  \BibitemOpen
  \bibfield  {author} {\bibinfo {author} {\bibfnamefont {N.}~\bibnamefont
  {Takeuchi}}, \bibinfo {author} {\bibfnamefont {Y.}~\bibnamefont
  {Yamanashi}},\ and\ \bibinfo {author} {\bibfnamefont {N.}~\bibnamefont
  {Yoshikawa}},\ }\bibfield  {title} {\bibinfo {title} {Reversibility and
  energy dissipation in adiabatic superconductor logic},\ }\href
  {https://doi.org/10.1038/s41598-017-00089-9} {\bibfield  {journal} {\bibinfo
  {journal} {Scientific Reports}\ }\textbf {\bibinfo {volume} {7}},\ \bibinfo
  {pages} {2045} (\bibinfo {year} {2017})}\BibitemShut {NoStop}%
\bibitem [{\citenamefont {Orlov}\ \emph {et~al.}(2012)\citenamefont {Orlov},
  \citenamefont {Lent}, \citenamefont {Thorpe}, \citenamefont {Boechler},\ and\
  \citenamefont {Snider}}]{Orlov_2012}%
  \BibitemOpen
  \bibfield  {author} {\bibinfo {author} {\bibfnamefont {A.~O.}\ \bibnamefont
  {Orlov}}, \bibinfo {author} {\bibfnamefont {C.~S.}\ \bibnamefont {Lent}},
  \bibinfo {author} {\bibfnamefont {C.~C.}\ \bibnamefont {Thorpe}}, \bibinfo
  {author} {\bibfnamefont {G.~P.}\ \bibnamefont {Boechler}},\ and\ \bibinfo
  {author} {\bibfnamefont {G.~L.}\ \bibnamefont {Snider}},\ }\bibfield  {title}
  {\bibinfo {title} {Experimental test of Landauer's principle
  at the
  sub-$k_B T$ level},\ }\href {https://doi.org/10.1143/jjap.51.06fe10}
  {\bibfield  {journal} {\bibinfo  {journal} {Japanese Journal of Applied
  Physics}\ }\textbf {\bibinfo {volume} {51}},\ \bibinfo {pages} {06FE10}
  (\bibinfo {year} {2012})}\BibitemShut {NoStop}%
\bibitem [{\citenamefont {Lent}\ \emph {et~al.}(2018)\citenamefont {Lent},
  \citenamefont {Orlov}, \citenamefont {Porod},\ and\ \citenamefont
  {Snider}}]{Snider_Book}%
  \BibitemOpen
  \bibinfo {editor} {\bibfnamefont {C.}~\bibnamefont {Lent}}, \bibinfo {editor}
  {\bibfnamefont {A.}~\bibnamefont {Orlov}}, \bibinfo {editor} {\bibfnamefont
  {W.}~\bibnamefont {Porod}},\ and\ \bibinfo {editor} {\bibfnamefont
  {G.}~\bibnamefont {Snider}},\ eds.,\ \href@noop {} {\emph {\bibinfo {title}
  {Energy Limits in Computation : A Review of }\\ {Landauer's Principle, Theory and
  Experiments}}}\ (\bibinfo  {publisher} {{Springer}},\ \bibinfo {year}
  {2018})\BibitemShut {NoStop}%
\bibitem [{\citenamefont {Borders}\ \emph {et~al.}(2019)\citenamefont
  {Borders}, \citenamefont {Pervaiz}, \citenamefont {Fukami}, \citenamefont
  {Camsari}, \citenamefont {Ohno},\ and\ \citenamefont {Datta}}]{Datta2019}%
  \BibitemOpen
  \bibfield  {author} {\bibinfo {author} {\bibfnamefont {W.~A.}\ \bibnamefont
  {Borders}}, \bibinfo {author} {\bibfnamefont {A.~Z.}\ \bibnamefont
  {Pervaiz}}, \bibinfo {author} {\bibfnamefont {S.}~\bibnamefont {Fukami}},
  \bibinfo {author} {\bibfnamefont {K.~Y.}\ \bibnamefont {Camsari}}, \bibinfo
  {author} {\bibfnamefont {H.}~\bibnamefont {Ohno}},\ and\ \bibinfo {author}
  {\bibfnamefont {S.}~\bibnamefont {Datta}},\ }\bibfield  {title} {\bibinfo
  {title} {Integer factorization using stochastic magnetic tunnel junctions},\
  }\href {https://doi.org/10.1038/s41586-019-1557-9} {\bibfield  {journal}
  {\bibinfo  {journal} {Nature}\ }\textbf {\bibinfo {volume} {573}},\ \bibinfo
  {pages} {390} (\bibinfo {year} {2019})}\BibitemShut {NoStop}%
\bibitem [{\citenamefont {Freitas}\ \emph {et~al.}(2021)\citenamefont
  {Freitas}, \citenamefont {Delvenne},\ and\ \citenamefont
  {Esposito}}]{Freitas2021}%
  \BibitemOpen
  \bibfield  {author} {\bibinfo {author} {\bibfnamefont {N.}~\bibnamefont
  {Freitas}}, \bibinfo {author} {\bibfnamefont {J.-C.}\ \bibnamefont
  {Delvenne}},\ and\ \bibinfo {author} {\bibfnamefont {M.}~\bibnamefont
  {Esposito}},\ }\bibfield  {title} {\bibinfo {title} {Stochastic
  thermodynamics of nonlinear electronic circuits: A realistic framework for
  computing around $k_B T$},\ }\href {https://doi.org/10.1103/PhysRevX.11.031064}
  {\bibfield  {journal} {\bibinfo  {journal} {Phys. Rev. X}\ }\textbf {\bibinfo
  {volume} {11}},\ \bibinfo {pages} {031064} (\bibinfo {year}
  {2021})}\BibitemShut {NoStop}%
\bibitem [{\citenamefont {Nielsen}\ and\ \citenamefont {Chuang}()}]{Chuang}%
  \BibitemOpen
  \bibinfo {editor} {\bibfnamefont {M.}~\bibnamefont {Nielsen}}\ and\ \bibinfo
  {editor} {\bibfnamefont {I.}~\bibnamefont {Chuang}},\ eds.,\ \href@noop {}
  {\emph {\bibinfo {title} {Quantum Computation and Quantum Information}}}\
  (\bibinfo  {publisher} {{Cambridge University Press}})\BibitemShut {NoStop}%
\bibitem [{\citenamefont {Haroche}\ and\ \citenamefont
  {Raimond}(1996)}]{QCnightmare}%
  \BibitemOpen
  \bibfield  {author} {\bibinfo {author} {\bibfnamefont {S.}~\bibnamefont
  {Haroche}}\ and\ \bibinfo {author} {\bibfnamefont {J.-M.}\ \bibnamefont
  {Raimond}},\ }\bibfield  {title} {\bibinfo {title} {Quantum computing: dream
  or nightmare},\ }\href@noop {} {\bibfield  {journal} {\bibinfo  {journal}
  {Physics Today}\ }\textbf {\bibinfo {volume} {49}},\ \bibinfo {pages} {51}
  (\bibinfo {year} {1996})}\BibitemShut {NoStop}%
\bibitem [{\citenamefont {Raussendorf}(2012)}]{Raussendorf}%
  \BibitemOpen
  \bibfield  {author} {\bibinfo {author} {\bibfnamefont {R.}~\bibnamefont
  {Raussendorf}},\ }\bibfield  {title} {\bibinfo {title} {{Key ideas in quantum
  error correction}},\ }\href {https://doi.org/10.1098/rsta.2011.0494}
  {\bibfield  {journal} {\bibinfo  {journal} {Philos. Trans. Royal Soc. A}\
  }\textbf {\bibinfo {volume} {370}},\ \bibinfo {pages} {4541} (\bibinfo {year}
  {2012})}\BibitemShut {NoStop}%
\bibitem [{\citenamefont {Devitt}\ \emph {et~al.}(2013)\citenamefont {Devitt},
  \citenamefont {Munro},\ and\ \citenamefont {Nemoto}}]{Devitt}%
  \BibitemOpen
  \bibfield  {author} {\bibinfo {author} {\bibfnamefont {S.~J.}\ \bibnamefont
  {Devitt}}, \bibinfo {author} {\bibfnamefont {W.~J.}\ \bibnamefont {Munro}},\
  and\ \bibinfo {author} {\bibfnamefont {K.}~\bibnamefont {Nemoto}},\
  }\bibfield  {title} {\bibinfo {title} {{Quantum error correction for
  beginners}},\ }\href {https://doi.org/10.1088/0034-4885/76/7/076001}
  {\bibfield  {journal} {\bibinfo  {journal} {Rep. Prog. Phys.}\ }\textbf
  {\bibinfo {volume} {76}},\ \bibinfo {pages} {076001} (\bibinfo {year}
  {2013})}\BibitemShut {NoStop}%
\bibitem [{\citenamefont {Gottesman}(2010)}]{Gottesman}%
  \BibitemOpen
  \bibfield  {author} {\bibinfo {author} {\bibfnamefont {D.}~\bibnamefont
  {Gottesman}},\ }\bibfield  {title} {\bibinfo {title} {{An introduction to
  quantum error correction and fault-tolerant quantum computation}},\ }in\
  \href@noop {} {\emph {\bibinfo {booktitle} {Proceedings of Symposia in
  Applied Mathematics}}},\ Vol.~\bibinfo {volume} {68},\ \bibinfo {editor}
  {edited by\ \bibinfo {editor} {\bibfnamefont {J.}~\bibnamefont {Samuel
  J.~Lomonaco}}}\ (\bibinfo  {publisher} {American Mathematical Society},\
  \bibinfo {year} {2010})\ pp.\ \bibinfo {pages} {13--58},\ \bibinfo {note}
  {{\href{https://arxiv.org/abs/0904.2557}{Eprint
  arXiv:0904.2557}}}\BibitemShut {NoStop}%
\bibitem [{\citenamefont {Zhong}(2020)}]{Pan2020}%
  \BibitemOpen
  \bibfield  {author} {\bibinfo {author} {\bibfnamefont {H.-S.}\ \bibnamefont
  {Zhong}},\ }\bibfield  {title} {\bibinfo {title} {Quantum computational
  advantage using photons},\ }\href {https://doi.org/10.1126/science.abe8770}
  {\ \textbf {\bibinfo {volume} {370}},\ \bibinfo {pages} {1460} (\bibinfo
  {year} {2020})}\BibitemShut {NoStop}%
\bibitem [{\citenamefont {Preskill}(2018)}]{NISQ}%
  \BibitemOpen
  \bibfield  {author} {\bibinfo {author} {\bibfnamefont {J.}~\bibnamefont
  {Preskill}},\ }\bibfield  {title} {\bibinfo {title} {Quantum {C}omputing in
  the {NISQ} era and beyond},\ }\href
  {https://doi.org/10.22331/q-2018-08-06-79} {\bibfield  {journal} {\bibinfo
  {journal} {{Quantum}}\ }\textbf {\bibinfo {volume} {2}},\ \bibinfo {pages}
  {79} (\bibinfo {year} {2018})}\BibitemShut {NoStop}%
\bibitem [{\citenamefont {Bharti}\ \emph {et~al.}(2022)\citenamefont {Bharti},
  \citenamefont {Cervera-Lierta}, \citenamefont {Kyaw}, \citenamefont {Haug},
  \citenamefont {Alperin-Lea}, \citenamefont {Anand}, \citenamefont {Degroote},
  \citenamefont {Heimonen}, \citenamefont {Kottmann}, \citenamefont {Menke},
  \citenamefont {Mok}, \citenamefont {Sim}, \citenamefont {Kwek},\ and\
  \citenamefont {Aspuru-Guzik}}]{Kwek}%
  \BibitemOpen
  \bibfield  {author} {\bibinfo {author} {\bibfnamefont {K.}~\bibnamefont
  {Bharti}}, \bibinfo {author} {\bibfnamefont {A.}~\bibnamefont
  {Cervera-Lierta}}, \bibinfo {author} {\bibfnamefont {T.~H.}\ \bibnamefont
  {Kyaw}}, \bibinfo {author} {\bibfnamefont {T.}~\bibnamefont {Haug}}, \bibinfo
  {author} {\bibfnamefont {S.}~\bibnamefont {Alperin-Lea}}, \bibinfo {author}
  {\bibfnamefont {A.}~\bibnamefont {Anand}}, \bibinfo {author} {\bibfnamefont
  {M.}~\bibnamefont {Degroote}}, \bibinfo {author} {\bibfnamefont
  {H.}~\bibnamefont {Heimonen}}, \bibinfo {author} {\bibfnamefont {J.~S.}\
  \bibnamefont {Kottmann}}, \bibinfo {author} {\bibfnamefont {T.}~\bibnamefont
  {Menke}}, \bibinfo {author} {\bibfnamefont {W.-K.}\ \bibnamefont {Mok}},
  \bibinfo {author} {\bibfnamefont {S.}~\bibnamefont {Sim}}, \bibinfo {author}
  {\bibfnamefont {L.-C.}\ \bibnamefont {Kwek}},\ and\ \bibinfo {author}
  {\bibfnamefont {A.}~\bibnamefont {Aspuru-Guzik}},\ }\bibfield  {title}
  {\bibinfo {title} {Noisy intermediate-scale quantum algorithms},\ }\href
  {https://doi.org/10.1103/RevModPhys.94.015004} {\bibfield  {journal}
  {\bibinfo  {journal} {Rev. Mod. Phys.}\ }\textbf {\bibinfo {volume} {94}},\
  \bibinfo {pages} {015004} (\bibinfo {year} {2022})}\BibitemShut {NoStop}%
\bibitem [{\citenamefont {Albash}\ and\ \citenamefont {Lidar}(2018)}]{AQC}%
  \BibitemOpen
  \bibfield  {author} {\bibinfo {author} {\bibfnamefont {T.}~\bibnamefont
  {Albash}}\ and\ \bibinfo {author} {\bibfnamefont {D.~A.}\ \bibnamefont
  {Lidar}},\ }\bibfield  {title} {\bibinfo {title} {Adiabatic quantum
  computation},\ }\href {https://doi.org/10.1103/RevModPhys.90.015002}
  {\bibfield  {journal} {\bibinfo  {journal} {Rev. Mod. Phys.}\ }\textbf
  {\bibinfo {volume} {90}},\ \bibinfo {pages} {015002} (\bibinfo {year}
  {2018})}\BibitemShut {NoStop}%
\bibitem [{\citenamefont {Gardiner}\ and\ \citenamefont {Zoller}()}]{noise}%
  \BibitemOpen
  \bibinfo {editor} {\bibfnamefont {C.}~\bibnamefont {Gardiner}}\ and\ \bibinfo
  {editor} {\bibfnamefont {P.}~\bibnamefont {Zoller}},\ eds.,\ \href@noop {}
  {\emph {\bibinfo {title} {Quantum Noise}}}\ (\bibinfo  {publisher}
  {{Springer}})\BibitemShut {NoStop}%
\bibitem [{\citenamefont {Deffner}\ and\ \citenamefont {Campbell}(2017)}]{QSL}%
  \BibitemOpen
  \bibfield  {author} {\bibinfo {author} {\bibfnamefont {S.}~\bibnamefont
  {Deffner}}\ and\ \bibinfo {author} {\bibfnamefont {S.}~\bibnamefont
  {Campbell}},\ }\bibfield  {title} {\bibinfo {title} {Quantum speed limits:
  from Heisenberg's uncertainty principle to optimal quantum control},\ }\href
  {https://doi.org/10.1088/1751-8121/aa86c6} {\bibfield  {journal} {\bibinfo
  {journal} {Journal of Physics A: Mathematical and Theoretical}\ }\textbf
  {\bibinfo {volume} {50}},\ \bibinfo {pages} {453001} (\bibinfo {year}
  {2017})}\BibitemShut {NoStop}%
\bibitem [{\citenamefont {Gu\'ery-Odelin}\ \emph {et~al.}(2019)\citenamefont
  {Gu\'ery-Odelin}, \citenamefont {Ruschhaupt}, \citenamefont {Kiely},
  \citenamefont {Torrontegui}, \citenamefont {Mart\'{\i}nez-Garaot},\ and\
  \citenamefont {Muga}}]{STA}%
  \BibitemOpen
  \bibfield  {author} {\bibinfo {author} {\bibfnamefont {D.}~\bibnamefont
  {Gu\'ery-Odelin}}, \bibinfo {author} {\bibfnamefont {A.}~\bibnamefont
  {Ruschhaupt}}, \bibinfo {author} {\bibfnamefont {A.}~\bibnamefont {Kiely}},
  \bibinfo {author} {\bibfnamefont {E.}~\bibnamefont {Torrontegui}}, \bibinfo
  {author} {\bibfnamefont {S.}~\bibnamefont {Mart\'{\i}nez-Garaot}},\ and\
  \bibinfo {author} {\bibfnamefont {J.~G.}\ \bibnamefont {Muga}},\ }\bibfield
  {title} {\bibinfo {title} {Shortcuts to adiabaticity: Concepts, methods, and
  applications},\ }\href {https://doi.org/10.1103/RevModPhys.91.045001}
  {\bibfield  {journal} {\bibinfo  {journal} {Rev. Mod. Phys.}\ }\textbf
  {\bibinfo {volume} {91}},\ \bibinfo {pages} {045001} (\bibinfo {year}
  {2019})}\BibitemShut {NoStop}%
\bibitem [{\citenamefont {Poyatos}\ \emph {et~al.}(1996)\citenamefont
  {Poyatos}, \citenamefont {Cirac},\ and\ \citenamefont {Zoller}}]{Zoller}%
  \BibitemOpen
  \bibfield  {author} {\bibinfo {author} {\bibfnamefont {J.~F.}\ \bibnamefont
  {Poyatos}}, \bibinfo {author} {\bibfnamefont {J.~I.}\ \bibnamefont {Cirac}},\
  and\ \bibinfo {author} {\bibfnamefont {P.}~\bibnamefont {Zoller}},\
  }\bibfield  {title} {\bibinfo {title} {Quantum reservoir engineering with
  laser cooled trapped ions},\ }\href
  {https://doi.org/10.1103/PhysRevLett.77.4728} {\bibfield  {journal} {\bibinfo
   {journal} {Phys. Rev. Lett.}\ }\textbf {\bibinfo {volume} {77}},\ \bibinfo
  {pages} {4728} (\bibinfo {year} {1996})}\BibitemShut {NoStop}%
\bibitem [{\citenamefont {Mirrahimi}(2016)}]{Mazyar}%
  \BibitemOpen
  \bibfield  {author} {\bibinfo {author} {\bibfnamefont {M.}~\bibnamefont
  {Mirrahimi}},\ }\bibfield  {title} {\bibinfo {title} {Cat-qubits for quantum
  computation},\ }\href
  {https://doi.org/https://doi.org/10.1016/j.crhy.2016.07.011} {\bibfield
  {journal} {\bibinfo  {journal} {Comptes Rendus Physique}\ }\textbf {\bibinfo
  {volume} {17}},\ \bibinfo {pages} {778} (\bibinfo {year} {2016})}\ \BibitemShut {NoStop}%
\bibitem [{\citenamefont {Lebreuilly}\ \emph {et~al.}(2021)\citenamefont
  {Lebreuilly}, \citenamefont {Noh}, \citenamefont {Wang}, \citenamefont
  {Girvin},\ and\ \citenamefont {Jiang}}]{Girvin}%
  \BibitemOpen
  \bibfield  {author} {\bibinfo {author} {\bibfnamefont {J.}~\bibnamefont
  {Lebreuilly}}, \bibinfo {author} {\bibfnamefont {K.}~\bibnamefont {Noh}},
  \bibinfo {author} {\bibfnamefont {C.-H.}\ \bibnamefont {Wang}}, \bibinfo
  {author} {\bibfnamefont {S.~M.}\ \bibnamefont {Girvin}},\ and\ \bibinfo
  {author} {\bibfnamefont {L.}~\bibnamefont {Jiang}},\ }\bibfield  {title}
  {\bibinfo {title} {Autonomous quantum error correction and quantum
  computation},\ }\bibfield  {journal} {\bibinfo  {journal} {arXiv}\ }\textbf
  {\bibinfo {volume} {2103.05007}},\ \href
  {https://doi.org/10.48550/arXiv.2103.05007} {10.48550/arXiv.2103.05007}
  (\bibinfo {year} {2021})\BibitemShut {NoStop}%
\bibitem [{\citenamefont {Vinjanampathy}\ and\ \citenamefont
  {Anders}(2016)}]{Janet-Sai}%
  \BibitemOpen
  \bibfield  {author} {\bibinfo {author} {\bibfnamefont {S.}~\bibnamefont
  {Vinjanampathy}}\ and\ \bibinfo {author} {\bibfnamefont {J.}~\bibnamefont
  {Anders}},\ }\bibfield  {title} {\bibinfo {title} {Quantum thermodynamics},\
  }\href {https://doi.org/10.1080/00107514.2016.1201896} {\bibfield  {journal}
  {\bibinfo  {journal} {Contemporary Physics}\ }\textbf {\bibinfo {volume}
  {57}},\ \bibinfo {pages} {545} (\bibinfo {year} {2016})},\ \Eprint
  {https://arxiv.org/abs/https://doi.org/10.1080/00107514.2016.1201896}
  {https://doi.org/10.1080/00107514.2016.1201896} \BibitemShut {NoStop}%
\bibitem [{\citenamefont {Binder}\ \emph {et~al.}(2018)\citenamefont {Binder},
  \citenamefont {Correa}, \citenamefont {Gogolin}, \citenamefont {anders},\
  and\ \citenamefont {Adesso}}]{Binder_book}%
  \BibitemOpen
  \bibinfo {editor} {\bibfnamefont {F.}~\bibnamefont {Binder}}, \bibinfo
  {editor} {\bibfnamefont {L.~A.}\ \bibnamefont {Correa}}, \bibinfo {editor}
  {\bibfnamefont {C.}~\bibnamefont {Gogolin}}, \bibinfo {editor} {\bibfnamefont
  {J.}~\bibnamefont {Anders}},\ and\ \bibinfo {editor} {\bibfnamefont
  {G.}~\bibnamefont {Adesso}},\ eds.,\ \href@noop {} {\emph {\bibinfo {title}
  {Thermodynamics in the {{Quantum Regime}}: }\\ {{{Fundamental Aspects}} and {{New
  Directions}}}}},\ Fundamental {{Theories}} of {{Physics}}\ (\bibinfo
  {publisher} {{Springer}},\ \bibinfo {address} {{Cham}},\ \bibinfo {year}
  {2018})\BibitemShut {NoStop}%
\bibitem [{\citenamefont {Chitambar}\ and\ \citenamefont
  {Gour}(2019)}]{RMP2019}%
  \BibitemOpen
  \bibfield  {author} {\bibinfo {author} {\bibfnamefont {E.}~\bibnamefont
  {Chitambar}}\ and\ \bibinfo {author} {\bibfnamefont {G.}~\bibnamefont
  {Gour}},\ }\bibfield  {title} {\bibinfo {title} {Quantum resource theories},\
  }\href {https://doi.org/10.1103/RevModPhys.91.025001} {\bibfield  {journal}
  {\bibinfo  {journal} {Rev. Mod. Phys.}\ }\textbf {\bibinfo {volume} {91}},\
  \bibinfo {pages} {025001} (\bibinfo {year} {2019})}\BibitemShut {NoStop}%
\bibitem [{\citenamefont {Korzekwa}\ \emph {et~al.}(2016)\citenamefont
  {Korzekwa}, \citenamefont {Lostaglio}, \citenamefont {Oppenheim},\ and\
  \citenamefont {Jennings}}]{korzekwaExtractionWorkQuantum2016}%
  \BibitemOpen
  \bibfield  {author} {\bibinfo {author} {\bibfnamefont {K.}~\bibnamefont
  {Korzekwa}}, \bibinfo {author} {\bibfnamefont {M.}~\bibnamefont {Lostaglio}},
  \bibinfo {author} {\bibfnamefont {J.}~\bibnamefont {Oppenheim}},\ and\
  \bibinfo {author} {\bibfnamefont {D.}~\bibnamefont {Jennings}},\ }\bibfield
  {title} {\bibinfo {title} {{The Extraction of Work from Quantum Coherence}},\
  }\href {https://doi.org/10.1088/1367-2630/18/2/023045} {\bibfield  {journal}
  {\bibinfo  {journal} {New J. Phys.}\ }\textbf {\bibinfo {volume} {18}},\
  \bibinfo {pages} {23045} (\bibinfo {year} {2016})}\BibitemShut {NoStop}%
\bibitem [{\citenamefont {Kammerlander}\ and\ \citenamefont
  {Anders}(2016)}]{Kammerlander2016}%
  \BibitemOpen
  \bibfield  {author} {\bibinfo {author} {\bibfnamefont {P.}~\bibnamefont
  {Kammerlander}}\ and\ \bibinfo {author} {\bibfnamefont {J.}~\bibnamefont
  {Anders}},\ }\bibfield  {title} {\bibinfo {title} {{Coherence and measurement
  in quantum thermodynamics}},\ }\href {https://doi.org/10.1038/srep22174}
  {\bibfield  {journal} {\bibinfo  {journal} {Sci. Rep.}\ }\textbf {\bibinfo
  {volume} {6}},\ \bibinfo {pages} {1} (\bibinfo {year} {2016})},\ \Eprint
  {https://arxiv.org/abs/1502.02673} {arXiv:1502.02673} \BibitemShut {NoStop}%
\bibitem [{\citenamefont {Lostaglio}\ \emph {et~al.}(2015)\citenamefont
  {Lostaglio}, \citenamefont {Jennings},\ and\ \citenamefont
  {Rudolph}}]{Lostaglio2015}%
  \BibitemOpen
  \bibfield  {author} {\bibinfo {author} {\bibfnamefont {M.}~\bibnamefont
  {Lostaglio}}, \bibinfo {author} {\bibfnamefont {D.}~\bibnamefont
  {Jennings}},\ and\ \bibinfo {author} {\bibfnamefont {T.}~\bibnamefont
  {Rudolph}},\ }\bibfield  {title} {\bibinfo {title} {{Description of quantum
  coherence in thermodynamic processes requires constraints beyond free
  energy}},\ }\href {https://doi.org/10.1038/ncomms7383} {\bibfield  {journal}
  {\bibinfo  {journal} {Nat. Commun.}\ }\textbf {\bibinfo {volume} {6}},\
  \bibinfo {pages} {1} (\bibinfo {year} {2015})},\ \Eprint
  {https://arxiv.org/abs/1405.2188} {arXiv:1405.2188} \BibitemShut {NoStop}%
\bibitem [{\citenamefont {Streltsov}\ \emph {et~al.}(2017)\citenamefont
  {Streltsov}, \citenamefont {Adesso},\ and\ \citenamefont
  {Plenio}}]{RMP_Coherence}%
  \BibitemOpen
  \bibfield  {author} {\bibinfo {author} {\bibfnamefont {A.}~\bibnamefont
  {Streltsov}}, \bibinfo {author} {\bibfnamefont {G.}~\bibnamefont {Adesso}},\
  and\ \bibinfo {author} {\bibfnamefont {M.~B.}\ \bibnamefont {Plenio}},\
  }\bibfield  {title} {\bibinfo {title} {Colloquium: Quantum coherence as a
  resource},\ }\href {https://doi.org/10.1103/RevModPhys.89.041003} {\bibfield
  {journal} {\bibinfo  {journal} {Rev. Mod. Phys.}\ }\textbf {\bibinfo {volume}
  {89}},\ \bibinfo {pages} {041003} (\bibinfo {year} {2017})}\BibitemShut
  {NoStop}%
\bibitem [{\citenamefont {Scovil}\ and\ \citenamefont
  {Schulz-DuBois}(1959)}]{Scovil59}%
  \BibitemOpen
  \bibfield  {author} {\bibinfo {author} {\bibfnamefont {H.~E.~D.}\
  \bibnamefont {Scovil}}\ and\ \bibinfo {author} {\bibfnamefont {E.~O.}\
  \bibnamefont {Schulz-DuBois}},\ }\bibfield  {title} {\bibinfo {title}
  {Three-level masers as heat engines},\ }\href
  {https://doi.org/10.1103/PhysRevLett.2.262} {\bibfield  {journal} {\bibinfo
  {journal} {Phys. Rev. Lett.}\ }\textbf {\bibinfo {volume} {2}},\ \bibinfo
  {pages} {262} (\bibinfo {year} {1959})}\BibitemShut {NoStop}%
\bibitem [{\citenamefont {Alicki}(1979)}]{Alicki1979}%
  \BibitemOpen
  \bibfield  {author} {\bibinfo {author} {\bibfnamefont {R.}~\bibnamefont
  {Alicki}},\ }\bibfield  {title} {\bibinfo {title} {{A quantum mechanical open
  system as a model of a heat engine}},\ }\bibfield  {journal} {\bibinfo
  {journal} {J. Chem. Phys.}\ }\textbf {\bibinfo {volume} {12}},\ \href
  {https://doi.org/10.1063/1.446862} {10.1063/1.446862} (\bibinfo {year}
  {1979})\BibitemShut {NoStop}%
\bibitem [{\citenamefont {Scully}\ \emph {et~al.}(2003)\citenamefont {Scully},
  \citenamefont {{Suhail Zubairy}}, \citenamefont {Agarwal},\ and\
  \citenamefont {Walther}}]{Scully2003}%
  \BibitemOpen
  \bibfield  {author} {\bibinfo {author} {\bibfnamefont {M.~O.}\ \bibnamefont
  {Scully}}, \bibinfo {author} {\bibfnamefont {M.}~\bibnamefont {{Suhail
  Zubairy}}}, \bibinfo {author} {\bibfnamefont {G.~S.}\ \bibnamefont
  {Agarwal}},\ and\ \bibinfo {author} {\bibfnamefont {H.}~\bibnamefont
  {Walther}},\ }\bibfield  {title} {\bibinfo {title} {{Extracting work from a
  single heat bath via vanishing quantum coherence}},\ }\href
  {https://doi.org/10.1126/science.1078955} {\bibfield  {journal} {\bibinfo
  {journal} {Science}\ }\textbf {\bibinfo {volume} {299}},\ \bibinfo {pages}
  {862} (\bibinfo {year} {2003})}\BibitemShut {NoStop}%
\bibitem [{\citenamefont {Uzdin}\ \emph {et~al.}(2015)\citenamefont {Uzdin},
  \citenamefont {Levy},\ and\ \citenamefont {Kosloff}}]{Uzdin2015}%
  \BibitemOpen
  \bibfield  {author} {\bibinfo {author} {\bibfnamefont {R.}~\bibnamefont
  {Uzdin}}, \bibinfo {author} {\bibfnamefont {A.}~\bibnamefont {Levy}},\ and\
  \bibinfo {author} {\bibfnamefont {R.}~\bibnamefont {Kosloff}},\ }\bibfield
  {title} {\bibinfo {title} {{Equivalence of quantum heat machines, and
  quantum-thermodynamic signatures}},\ }\href
  {https://doi.org/10.1103/PhysRevX.5.031044} {\bibfield  {journal} {\bibinfo
  {journal} {Phys. Rev. X}\ }\textbf {\bibinfo {volume} {5}},\ \bibinfo {pages}
  {1} (\bibinfo {year} {2015})}\BibitemShut {NoStop}%
\bibitem [{\citenamefont {Niedenzu}\ \emph {et~al.}(2018)\citenamefont
  {Niedenzu}, \citenamefont {Mukherjee}, \citenamefont {Ghosh}, \citenamefont
  {Kofman},\ and\ \citenamefont {Kurizki}}]{Kurizki}%
  \BibitemOpen
  \bibfield  {author} {\bibinfo {author} {\bibfnamefont {W.}~\bibnamefont
  {Niedenzu}}, \bibinfo {author} {\bibfnamefont {V.}~\bibnamefont {Mukherjee}},
  \bibinfo {author} {\bibfnamefont {A.}~\bibnamefont {Ghosh}}, \bibinfo
  {author} {\bibfnamefont {A.}~\bibnamefont {Kofman}},\ and\ \bibinfo {author}
  {\bibfnamefont {G.}~\bibnamefont {Kurizki}},\ }\bibfield  {title} {\bibinfo
  {title} {Quantum engine efficiency bound beyond the second law of
  thermodynamics},\ }\href {https://doi.org/10.1038/s41467-017-01991-6}
  {\bibfield  {journal} {\bibinfo  {journal} {Nat Commun.}\ }\textbf {\bibinfo
  {volume} {9}},\ \bibinfo {pages} {165} (\bibinfo {year} {2018})}\BibitemShut
  {NoStop}%
\bibitem [{\citenamefont {Levy}\ \emph {et~al.}(2016)\citenamefont {Levy},
  \citenamefont {Di\'osi},\ and\ \citenamefont {Kosloff}}]{Ronnie2016}%
  \BibitemOpen
  \bibfield  {author} {\bibinfo {author} {\bibfnamefont {A.}~\bibnamefont
  {Levy}}, \bibinfo {author} {\bibfnamefont {L.}~\bibnamefont {Di\'osi}},\ and\
  \bibinfo {author} {\bibfnamefont {R.}~\bibnamefont {Kosloff}},\ }\bibfield
  {title} {\bibinfo {title} {Quantum flywheel},\ }\href
  {https://doi.org/10.1103/PhysRevA.93.052119} {\bibfield  {journal} {\bibinfo
  {journal} {Phys. Rev. A}\ }\textbf {\bibinfo {volume} {93}},\ \bibinfo
  {pages} {052119} (\bibinfo {year} {2016})}\BibitemShut {NoStop}%
\bibitem [{\citenamefont {Goold}\ \emph {et~al.}(2016)\citenamefont {Goold},
  \citenamefont {Huber}, \citenamefont {Riera}, \citenamefont {del Rio},\ and\
  \citenamefont {Skrzypczyk}}]{Goold_2016}%
  \BibitemOpen
  \bibfield  {author} {\bibinfo {author} {\bibfnamefont {J.}~\bibnamefont
  {Goold}}, \bibinfo {author} {\bibfnamefont {M.}~\bibnamefont {Huber}},
  \bibinfo {author} {\bibfnamefont {A.}~\bibnamefont {Riera}}, \bibinfo
  {author} {\bibfnamefont {L.}~\bibnamefont {del Rio}},\ and\ \bibinfo {author}
  {\bibfnamefont {P.}~\bibnamefont {Skrzypczyk}},\ }\bibfield  {title}
  {\bibinfo {title} {The role of quantum information in
  thermodynamics{\textemdash}a topical review},\ }\href
  {https://doi.org/10.1088/1751-8113/49/14/143001} {\bibfield  {journal}
  {\bibinfo  {journal} {Journal of Physics A: Mathematical and Theoretical}\
  }\textbf {\bibinfo {volume} {49}},\ \bibinfo {pages} {143001} (\bibinfo
  {year} {2016})}\BibitemShut {NoStop}%
\bibitem [{\citenamefont {Goold}\ \emph {et~al.}(2015)\citenamefont {Goold},
  \citenamefont {Paternostro},\ and\ \citenamefont {Modi}}]{Quantum_Landauer}%
  \BibitemOpen
  \bibfield  {author} {\bibinfo {author} {\bibfnamefont {J.}~\bibnamefont
  {Goold}}, \bibinfo {author} {\bibfnamefont {M.}~\bibnamefont {Paternostro}},\
  and\ \bibinfo {author} {\bibfnamefont {K.}~\bibnamefont {Modi}},\ }\bibfield
  {title} {\bibinfo {title} {Nonequilibrium quantum Landauer principle},\
  }\href {https://doi.org/10.1103/PhysRevLett.114.060602} {\bibfield  {journal}
  {\bibinfo  {journal} {Phys. Rev. Lett.}\ }\textbf {\bibinfo {volume} {114}},\
  \bibinfo {pages} {060602} (\bibinfo {year} {2015})}\BibitemShut {NoStop}%
\bibitem [{\citenamefont {Bedingham}\ and\ \citenamefont
  {Maroney}(2016)}]{Owen}%
  \BibitemOpen
  \bibfield  {author} {\bibinfo {author} {\bibfnamefont {D.~J.}\ \bibnamefont
  {Bedingham}}\ and\ \bibinfo {author} {\bibfnamefont {O.~J.~E.}\ \bibnamefont
  {Maroney}},\ }\bibfield  {title} {\bibinfo {title} {The thermodynamic cost of
  quantum operations},\ }\href {https://doi.org/10.1088/1367-2630/18/11/113050}
  {\bibfield  {journal} {\bibinfo  {journal} {New Journal of Physics}\ }\textbf
  {\bibinfo {volume} {18}},\ \bibinfo {pages} {113050} (\bibinfo {year}
  {2016})}\BibitemShut {NoStop}%
\bibitem [{\citenamefont {Cottet}\ \emph {et~al.}(2017)\citenamefont {Cottet},
  \citenamefont {Jezouin}, \citenamefont {Bretheau}, \citenamefont
  {Campagne-Ibarcq}, \citenamefont {Ficheux}, \citenamefont {anders},
  \citenamefont {Auff{\`e}ves}, \citenamefont {Azouit}, \citenamefont
  {Rouchon},\ and\ \citenamefont {Huard}}]{Huard}%
  \BibitemOpen
  \bibfield  {author} {\bibinfo {author} {\bibfnamefont {N.}~\bibnamefont
  {Cottet}}, \bibinfo {author} {\bibfnamefont {S.}~\bibnamefont {Jezouin}},
  \bibinfo {author} {\bibfnamefont {L.}~\bibnamefont {Bretheau}}, \bibinfo
  {author} {\bibfnamefont {P.}~\bibnamefont {Campagne-Ibarcq}}, \bibinfo
  {author} {\bibfnamefont {Q.}~\bibnamefont {Ficheux}}, \bibinfo {author}
  {\bibfnamefont {J.}~\bibnamefont {anders}}, \bibinfo {author} {\bibfnamefont
  {A.}~\bibnamefont {Auff{\`e}ves}}, \bibinfo {author} {\bibfnamefont
  {R.}~\bibnamefont {Azouit}}, \bibinfo {author} {\bibfnamefont
  {P.}~\bibnamefont {Rouchon}},\ and\ \bibinfo {author} {\bibfnamefont
  {B.}~\bibnamefont {Huard}},\ }\bibfield  {title} {\bibinfo {title} {Observing
  a quantum Maxwell demon at work},\ }\href@noop {} {\bibfield  {journal}
  {\bibinfo  {journal} {Proceedings of the National Academy of Sciences}\
  }\textbf {\bibinfo {volume} {114}},\ \bibinfo {pages} {7561} (\bibinfo {year}
  {2017})}\BibitemShut {NoStop}%
\bibitem [{\citenamefont {Naghiloo}\ \emph {et~al.}(2018)\citenamefont
  {Naghiloo}, \citenamefont {Alonso}, \citenamefont {Romito}, \citenamefont
  {Lutz},\ and\ \citenamefont {Murch}}]{Murch}%
  \BibitemOpen
  \bibfield  {author} {\bibinfo {author} {\bibfnamefont {M.}~\bibnamefont
  {Naghiloo}}, \bibinfo {author} {\bibfnamefont {J.~J.}\ \bibnamefont
  {Alonso}}, \bibinfo {author} {\bibfnamefont {A.}~\bibnamefont {Romito}},
  \bibinfo {author} {\bibfnamefont {E.}~\bibnamefont {Lutz}},\ and\ \bibinfo
  {author} {\bibfnamefont {K.~W.}\ \bibnamefont {Murch}},\ }\bibfield  {title}
  {\bibinfo {title} {Information gain and loss for a quantum Maxwell's demon},\
  }\href {https://doi.org/10.1103/PhysRevLett.121.030604} {\bibfield  {journal}
  {\bibinfo  {journal} {Phys. Rev. Lett.}\ }\textbf {\bibinfo {volume} {121}},\
  \bibinfo {pages} {030604} (\bibinfo {year} {2018})}\BibitemShut {NoStop}%
\bibitem [{\citenamefont {Masuyama}\ \emph {et~al.}(2018)\citenamefont
  {Masuyama}, \citenamefont {Funo}, \citenamefont {Murashita}, \citenamefont
  {Noguchi}, \citenamefont {Kono}, \citenamefont {Tabuchi}, \citenamefont
  {Yamazaki}, \citenamefont {Ueda},\ and\ \citenamefont
  {Nakamura}}]{Masuyama18}%
  \BibitemOpen
  \bibfield  {author} {\bibinfo {author} {\bibfnamefont {Y.}~\bibnamefont
  {Masuyama}}, \bibinfo {author} {\bibfnamefont {K.}~\bibnamefont {Funo}},
  \bibinfo {author} {\bibfnamefont {Y.}~\bibnamefont {Murashita}}, \bibinfo
  {author} {\bibfnamefont {A.}~\bibnamefont {Noguchi}}, \bibinfo {author}
  {\bibfnamefont {S.}~\bibnamefont {Kono}}, \bibinfo {author} {\bibfnamefont
  {Y.}~\bibnamefont {Tabuchi}}, \bibinfo {author} {\bibfnamefont
  {R.}~\bibnamefont {Yamazaki}}, \bibinfo {author} {\bibfnamefont
  {M.}~\bibnamefont {Ueda}},\ and\ \bibinfo {author} {\bibfnamefont
  {Y.}~\bibnamefont {Nakamura}},\ }\bibfield  {title} {\bibinfo {title}
  {Information-to-work conversion by Maxwell's demon in a superconducting
  circuit quantum electrodynamical system},\ }\href
  {https://doi.org/10.1038/s41467-018-03686-y} {\bibfield  {journal} {\bibinfo
  {journal} {{Nature Communications}}\ }\textbf {\bibinfo {volume} {9}},\
  \bibinfo {pages} {1291} (\bibinfo {year} {2018})}\BibitemShut {NoStop}%
\bibitem [{\citenamefont {Strasberg}(2022)}]{Strasberg_Book}%
  \BibitemOpen
  \bibfield  {author} {\bibinfo {author} {\bibfnamefont {P.}~\bibnamefont
  {Strasberg}},\ }\href@noop {} {\emph {\bibinfo {title} {Quantum Stochastic
  Thermodynamics: }\\ {Foundations and Selected Applications}}}\ (\bibinfo
  {publisher} {{Oxford Graduate Texts}},\ \bibinfo {year} {2022})\BibitemShut
  {NoStop}%
\bibitem [{\citenamefont {Campisi}\ \emph {et~al.}(2011)\citenamefont
  {Campisi}, \citenamefont {H\"anggi},\ and\ \citenamefont {Talkner}}]{Hanggi}%
  \BibitemOpen
  \bibfield  {author} {\bibinfo {author} {\bibfnamefont {M.}~\bibnamefont
  {Campisi}}, \bibinfo {author} {\bibfnamefont {P.}~\bibnamefont {H\"anggi}},\
  and\ \bibinfo {author} {\bibfnamefont {P.}~\bibnamefont {Talkner}},\
  }\bibfield  {title} {\bibinfo {title} {Colloquium: Quantum fluctuation
  relations: Foundations and applications},\ }\href
  {https://doi.org/10.1103/RevModPhys.83.771} {\bibfield  {journal} {\bibinfo
  {journal} {Rev. Mod. Phys.}\ }\textbf {\bibinfo {volume} {83}},\ \bibinfo
  {pages} {771} (\bibinfo {year} {2011})}\BibitemShut {NoStop}%
\bibitem [{\citenamefont {Funo}\ \emph {et~al.}(2013)\citenamefont {Funo},
  \citenamefont {Watanabe},\ and\ \citenamefont {Ueda}}]{Funo}%
  \BibitemOpen
  \bibfield  {author} {\bibinfo {author} {\bibfnamefont {K.}~\bibnamefont
  {Funo}}, \bibinfo {author} {\bibfnamefont {Y.}~\bibnamefont {Watanabe}},\
  and\ \bibinfo {author} {\bibfnamefont {M.}~\bibnamefont {Ueda}},\ }\bibfield
  {title} {\bibinfo {title} {Integral quantum fluctuation theorems under
  measurement and feedback control},\ }\href
  {https://doi.org/10.1103/PhysRevE.88.052121} {\bibfield  {journal} {\bibinfo
  {journal} {Phys. Rev. E}\ }\textbf {\bibinfo {volume} {88}},\ \bibinfo
  {pages} {052121} (\bibinfo {year} {2013})}\BibitemShut {NoStop}%
\bibitem [{\citenamefont {Horowitz}\ and\ \citenamefont
  {Gingrich}(2020)}]{TUR}%
  \BibitemOpen
  \bibfield  {author} {\bibinfo {author} {\bibfnamefont {J.~M.}\ \bibnamefont
  {Horowitz}}\ and\ \bibinfo {author} {\bibfnamefont {T.~R.}\ \bibnamefont
  {Gingrich}},\ }\bibfield  {title} {\bibinfo {title} {Thermodynamic
  uncertainty relations constrain non-equilibrium fluctuations},\ }\href
  {https://doi.org/10.1038/s41567-019-0702-6} {\bibfield  {journal} {\bibinfo
  {journal} {Nature Physics}\ }\textbf {\bibinfo {volume} {16}},\ \bibinfo
  {pages} {15} (\bibinfo {year} {2020})}\BibitemShut {NoStop}%
\bibitem [{\citenamefont {Landi}\ and\ \citenamefont
  {Paternostro}(2021)}]{Paternostro}%
  \BibitemOpen
  \bibfield  {author} {\bibinfo {author} {\bibfnamefont {G.~T.}\ \bibnamefont
  {Landi}}\ and\ \bibinfo {author} {\bibfnamefont {M.}~\bibnamefont
  {Paternostro}},\ }\bibfield  {title} {\bibinfo {title} {Irreversible entropy
  production: From classical to quantum},\ }\href
  {https://doi.org/10.1103/RevModPhys.93.035008} {\bibfield  {journal}
  {\bibinfo  {journal} {Rev. Mod. Phys.}\ }\textbf {\bibinfo {volume} {93}},\
  \bibinfo {pages} {035008} (\bibinfo {year} {2021})}\BibitemShut {NoStop}%
\bibitem [{\citenamefont {Gea-Banacloche}(2002)}]{Gea}%
  \BibitemOpen
  \bibfield  {author} {\bibinfo {author} {\bibfnamefont {J.}~\bibnamefont
  {Gea-Banacloche}},\ }\bibfield  {title} {\bibinfo {title} {Minimum energy
  requirements for quantum computation},\ }\href
  {https://doi.org/10.1103/PhysRevLett.89.217901} {\bibfield  {journal}
  {\bibinfo  {journal} {Phys. Rev. Lett.}\ }\textbf {\bibinfo {volume} {89}},\
  \bibinfo {pages} {217901} (\bibinfo {year} {2002})}\BibitemShut {NoStop}%
\bibitem [{\citenamefont {Ikonen}\ \emph {et~al.}(2017)\citenamefont {Ikonen},
  \citenamefont {Salmilehto},\ and\ \citenamefont
  {M{\"o}tt{\"o}nen}}]{Mottonen}%
  \BibitemOpen
  \bibfield  {author} {\bibinfo {author} {\bibfnamefont {J.}~\bibnamefont
  {Ikonen}}, \bibinfo {author} {\bibfnamefont {J.}~\bibnamefont {Salmilehto}},\
  and\ \bibinfo {author} {\bibfnamefont {M.}~\bibnamefont {M{\"o}tt{\"o}nen}},\
  }\bibfield  {title} {\bibinfo {title} {Energy-efficient quantum computing},\
  }\href {https://doi.org/10.1038/s41534-017-0015-5} {\bibfield  {journal}
  {\bibinfo  {journal} {npj Quantum Information}\ }\textbf {\bibinfo {volume}
  {3}},\ \bibinfo {pages} {17} (\bibinfo {year} {2017})}\BibitemShut {NoStop}%
\bibitem [{\citenamefont {Cimini}\ \emph {et~al.}(2020)\citenamefont {Cimini},
  \citenamefont {Gherardini}, \citenamefont {Barbieri}, \citenamefont
  {Gianadi}, \citenamefont {Sbroscia}, \citenamefont {Buffoni}, \citenamefont
  {Paternostro},\ and\ \citenamefont {Caruso}}]{Barbieri}%
  \BibitemOpen
  \bibfield  {author} {\bibinfo {author} {\bibfnamefont {V.}~\bibnamefont
  {Cimini}}, \bibinfo {author} {\bibfnamefont {S.}~\bibnamefont {Gherardini}},
  \bibinfo {author} {\bibfnamefont {M.}~\bibnamefont {Barbieri}}, \bibinfo
  {author} {\bibfnamefont {I.}~\bibnamefont {Gianadi}}, \bibinfo {author}
  {\bibfnamefont {M.}~\bibnamefont {Sbroscia}}, \bibinfo {author}
  {\bibfnamefont {L.}~\bibnamefont {Buffoni}}, \bibinfo {author} {\bibfnamefont
  {M.}~\bibnamefont {Paternostro}},\ and\ \bibinfo {author} {\bibfnamefont
  {F.}~\bibnamefont {Caruso}},\ }\bibfield  {title} {\bibinfo {title}
  {Experimental characterization of the energetics of quantum logic gates},\
  }\href {https://doi.org/10.1038/s41534-020-00325-7} {\bibfield  {journal}
  {\bibinfo  {journal} {npj Quantum Inf}\ }\textbf {\bibinfo {volume} {6}},\
  \bibinfo {pages} {96} (\bibinfo {year} {2020})}\BibitemShut {NoStop}%
\bibitem [{\citenamefont {Stevens}()}]{Ben21}%
  \BibitemOpen
  \bibfield  {author} {\bibinfo {author} {\bibfnamefont {J.}~\bibnamefont
  {Stevens}},\ }\bibfield  {title} {\bibinfo {title} {Energetics of a single
  qubit gate},\ }\href@noop {} {\bibfield  {journal} {\bibinfo  {journal}
  {arXiv}\ }\textbf {\bibinfo {volume} {2109.09648}}}\BibitemShut {NoStop}%
\bibitem [{\citenamefont {Deffner}(2021)}]{Deffner}%
  \BibitemOpen
  \bibfield  {author} {\bibinfo {author} {\bibfnamefont {S.}~\bibnamefont
  {Deffner}},\ }\bibfield  {title} {\bibinfo {title} {Energetic cost of
  hamiltonian quantum gates},\ }\href
  {https://doi.org/10.1209/0295-5075/134/40002} {\bibfield  {journal} {\bibinfo
   {journal} {EPL}\ }\textbf {\bibinfo {volume} {134}},\ \bibinfo {pages}
  {40002} (\bibinfo {year} {2021})}\BibitemShut {NoStop}%
\bibitem [{\citenamefont {Caves}(1982)}]{Caves82}%
  \BibitemOpen
  \bibfield  {author} {\bibinfo {author} {\bibfnamefont {C.~M.}\ \bibnamefont
  {Caves}},\ }\bibfield  {title} {\bibinfo {title} {Quantum limits on noise in
  linear amplifiers},\ }\href {https://doi.org/10.1103/PhysRevD.26.1817}
  {\bibfield  {journal} {\bibinfo  {journal} {Phys. Rev. D}\ }\textbf {\bibinfo
  {volume} {26}},\ \bibinfo {pages} {1817} (\bibinfo {year}
  {1982})}\BibitemShut {NoStop}%
\bibitem [{\citenamefont {Caves}\ and\ \citenamefont
  {Drummond}(1994)}]{Caves94}%
  \BibitemOpen
  \bibfield  {author} {\bibinfo {author} {\bibfnamefont {C.~M.}\ \bibnamefont
  {Caves}}\ and\ \bibinfo {author} {\bibfnamefont {P.~D.}\ \bibnamefont
  {Drummond}},\ }\bibfield  {title} {\bibinfo {title} {Quantum limits on
  bosonic communication rates},\ }\href
  {https://doi.org/10.1103/RevModPhys.66.481} {\bibfield  {journal} {\bibinfo
  {journal} {Rev. Mod. Phys.}\ }\textbf {\bibinfo {volume} {66}},\ \bibinfo
  {pages} {481} (\bibinfo {year} {1994})}\BibitemShut {NoStop}%
\bibitem [{\citenamefont {Guryanova}\ \emph {et~al.}(2020)\citenamefont
  {Guryanova}, \citenamefont {Friis},\ and\ \citenamefont {Huber}}]{Huber}%
  \BibitemOpen
  \bibfield  {author} {\bibinfo {author} {\bibfnamefont {Y.}~\bibnamefont
  {Guryanova}}, \bibinfo {author} {\bibfnamefont {N.}~\bibnamefont {Friis}},\
  and\ \bibinfo {author} {\bibfnamefont {M.}~\bibnamefont {Huber}},\ }\bibfield
   {title} {\bibinfo {title} {Ideal {P}rojective {M}easurements {H}ave
  {I}nfinite {R}esource {C}osts},\ }\href
  {https://doi.org/10.22331/q-2020-01-13-222} {\bibfield  {journal} {\bibinfo
  {journal} {{Quantum}}\ }\textbf {\bibinfo {volume} {4}},\ \bibinfo {pages}
  {222} (\bibinfo {year} {2020})}\BibitemShut {NoStop}%
\bibitem [{\citenamefont {Sartori}\ and\ \citenamefont
  {Pigolotti}(2015)}]{MD_QEC}%
  \BibitemOpen
  \bibfield  {author} {\bibinfo {author} {\bibfnamefont {P.}~\bibnamefont
  {Sartori}}\ and\ \bibinfo {author} {\bibfnamefont {S.}~\bibnamefont
  {Pigolotti}},\ }\bibfield  {title} {\bibinfo {title} {Thermodynamics of error
  correction},\ }\href {https://doi.org/10.1103/PhysRevX.5.041039} {\bibfield
  {journal} {\bibinfo  {journal} {Phys. Rev. X}\ }\textbf {\bibinfo {volume}
  {5}},\ \bibinfo {pages} {041039} (\bibinfo {year} {2015})}\BibitemShut
  {NoStop}%
\bibitem [{\citenamefont {Landi}\ \emph {et~al.}(2020)\citenamefont {Landi},
  \citenamefont {Fonseca~de Oliveira},\ and\ \citenamefont {Buksman}}]{Landi}%
  \BibitemOpen
  \bibfield  {author} {\bibinfo {author} {\bibfnamefont {G.~T.}\ \bibnamefont
  {Landi}}, \bibinfo {author} {\bibfnamefont {a.~L.}\ \bibnamefont {Fonseca~de
  Oliveira}},\ and\ \bibinfo {author} {\bibfnamefont {E.}~\bibnamefont
  {Buksman}},\ }\bibfield  {title} {\bibinfo {title} {Thermodynamic analysis of
  quantum error-correcting engines},\ }\href
  {https://doi.org/10.1103/PhysRevA.101.042106} {\bibfield  {journal} {\bibinfo
   {journal} {Phys. Rev. A}\ }\textbf {\bibinfo {volume} {101}},\ \bibinfo
  {pages} {042106} (\bibinfo {year} {2020})}\BibitemShut {NoStop}%
\bibitem [{\citenamefont {Buffoni}\ and\ \citenamefont
  {Campisi}(2020)}]{Campisi}%
  \BibitemOpen
  \bibfield  {author} {\bibinfo {author} {\bibfnamefont {L.}~\bibnamefont
  {Buffoni}}\ and\ \bibinfo {author} {\bibfnamefont {M.}~\bibnamefont
  {Campisi}},\ }\bibfield  {title} {\bibinfo {title} {Thermodynamics of a
  quantum annealer},\ }\href@noop {} {\bibfield  {journal} {\bibinfo  {journal}
  {Quantum Science and Technology}\ } (\bibinfo {year} {2020})}\BibitemShut
  {NoStop}%
\bibitem [{\citenamefont {Campisi}\ and\ \citenamefont
  {Buffoni}(2021)}]{Campisi21}%
  \BibitemOpen
  \bibfield  {author} {\bibinfo {author} {\bibfnamefont {M.}~\bibnamefont
  {Campisi}}\ and\ \bibinfo {author} {\bibfnamefont {L.}~\bibnamefont
  {Buffoni}},\ }\bibfield  {title} {\bibinfo {title} {Improved bound on entropy
  production in a quantum annealer},\ }\href
  {https://doi.org/10.1103/PhysRevE.104.L022102} {\bibfield  {journal}
  {\bibinfo  {journal} {Phys. Rev. E}\ }\textbf {\bibinfo {volume} {104}},\
  \bibinfo {pages} {L022102} (\bibinfo {year} {2021})}\BibitemShut {NoStop}%
\bibitem [{\citenamefont {Campaioli}\ \emph {et~al.}(2017)\citenamefont
  {Campaioli}, \citenamefont {Pollock}, \citenamefont {Binder}, \citenamefont
  {C{\'{e}}leri}, \citenamefont {Goold}, \citenamefont {Vinjanampathy},\ and\
  \citenamefont {Modi}}]{Campaioli2017}%
  \BibitemOpen
  \bibfield  {author} {\bibinfo {author} {\bibfnamefont {F.}~\bibnamefont
  {Campaioli}}, \bibinfo {author} {\bibfnamefont {F.~A.}\ \bibnamefont
  {Pollock}}, \bibinfo {author} {\bibfnamefont {F.~C.}\ \bibnamefont {Binder}},
  \bibinfo {author} {\bibfnamefont {L.}~\bibnamefont {C{\'{e}}leri}}, \bibinfo
  {author} {\bibfnamefont {J.}~\bibnamefont {Goold}}, \bibinfo {author}
  {\bibfnamefont {S.}~\bibnamefont {Vinjanampathy}},\ and\ \bibinfo {author}
  {\bibfnamefont {K.}~\bibnamefont {Modi}},\ }\bibfield  {title} {\bibinfo
  {title} {{Enhancing the Charging Power of Quantum Batteries}},\ }\href
  {https://doi.org/10.1103/PhysRevLett.118.150601} {\bibfield  {journal}
  {\bibinfo  {journal} {Phys. Rev. Lett.}\ }\textbf {\bibinfo {volume} {118}},\
  \bibinfo {pages} {1} (\bibinfo {year} {2017})},\ \Eprint
  {https://arxiv.org/abs/1612.04991} {arXiv:1612.04991} \BibitemShut {NoStop}%
\bibitem [{\citenamefont {Ferraro}\ \emph {et~al.}(2018)\citenamefont
  {Ferraro}, \citenamefont {Campisi}, \citenamefont {Andolina}, \citenamefont
  {Pellegrini},\ and\ \citenamefont {Polini}}]{Ferraro2018}%
  \BibitemOpen
  \bibfield  {author} {\bibinfo {author} {\bibfnamefont {D.}~\bibnamefont
  {Ferraro}}, \bibinfo {author} {\bibfnamefont {M.}~\bibnamefont {Campisi}},
  \bibinfo {author} {\bibfnamefont {G.~M.}\ \bibnamefont {Andolina}}, \bibinfo
  {author} {\bibfnamefont {V.}~\bibnamefont {Pellegrini}},\ and\ \bibinfo
  {author} {\bibfnamefont {M.}~\bibnamefont {Polini}},\ }\bibfield  {title}
  {\bibinfo {title} {{High-Power Collective Charging of a Solid-State Quantum
  Battery}},\ }\href {https://doi.org/10.1103/PhysRevLett.120.117702}
  {\bibfield  {journal} {\bibinfo  {journal} {Phys. Rev. Lett.}\ }\textbf
  {\bibinfo {volume} {120}},\ \bibinfo {pages} {117702} (\bibinfo {year}
  {2018})},\ \Eprint {https://arxiv.org/abs/1707.04930} {arXiv:1707.04930}
  \BibitemShut {NoStop}%
\bibitem [{\citenamefont {Binder}\ \emph {et~al.}(2015)\citenamefont {Binder},
  \citenamefont {Vinjanampathy}, \citenamefont {Modi},\ and\ \citenamefont
  {Goold}}]{Quantacell}%
  \BibitemOpen
  \bibfield  {author} {\bibinfo {author} {\bibfnamefont {F.~C.}\ \bibnamefont
  {Binder}}, \bibinfo {author} {\bibfnamefont {S.}~\bibnamefont
  {Vinjanampathy}}, \bibinfo {author} {\bibfnamefont {K.}~\bibnamefont
  {Modi}},\ and\ \bibinfo {author} {\bibfnamefont {J.}~\bibnamefont {Goold}},\
  }\bibfield  {title} {\bibinfo {title} {{Quantacell: Powerful charging of
  quantum batteries}},\ }\bibfield  {journal} {\bibinfo  {journal} {New J.
  Phys.}\ }\textbf {\bibinfo {volume} {17}},\ \href
  {https://doi.org/10.1088/1367-2630/17/7/075015}
  {10.1088/1367-2630/17/7/075015} (\bibinfo {year} {2015})\BibitemShut
  {NoStop}%
\bibitem [{\citenamefont {Andolina}\ \emph {et~al.}(2019)\citenamefont
  {Andolina}, \citenamefont {Keck}, \citenamefont {Mari}, \citenamefont
  {Campisi}, \citenamefont {Giovannetti},\ and\ \citenamefont
  {Polini}}]{Polini2019}%
  \BibitemOpen
  \bibfield  {author} {\bibinfo {author} {\bibfnamefont {G.~M.}\ \bibnamefont
  {Andolina}}, \bibinfo {author} {\bibfnamefont {M.}~\bibnamefont {Keck}},
  \bibinfo {author} {\bibfnamefont {A.}~\bibnamefont {Mari}}, \bibinfo {author}
  {\bibfnamefont {M.}~\bibnamefont {Campisi}}, \bibinfo {author} {\bibfnamefont
  {V.}~\bibnamefont {Giovannetti}},\ and\ \bibinfo {author} {\bibfnamefont
  {M.}~\bibnamefont {Polini}},\ }\bibfield  {title} {\bibinfo {title}
  {Extractable work, the role of correlations, and asymptotic freedom in
  quantum batteries},\ }\href {https://doi.org/10.1103/PhysRevLett.122.047702}
  {\bibfield  {journal} {\bibinfo  {journal} {Phys. Rev. Lett.}\ }\textbf
  {\bibinfo {volume} {122}},\ \bibinfo {pages} {047702} (\bibinfo {year}
  {2019})}\BibitemShut {NoStop}%
\bibitem [{\citenamefont {Juli{\`{a}}-Farr{\'{e}}}\ \emph
  {et~al.}(2020)\citenamefont {Juli{\`{a}}-Farr{\'{e}}}, \citenamefont
  {Salamon}, \citenamefont {Riera}, \citenamefont {Bera},\ and\ \citenamefont
  {Lewenstein}}]{Julia-Farre2020}%
  \BibitemOpen
  \bibfield  {author} {\bibinfo {author} {\bibfnamefont {S.}~\bibnamefont
  {Juli{\`{a}}-Farr{\'{e}}}}, \bibinfo {author} {\bibfnamefont
  {T.}~\bibnamefont {Salamon}}, \bibinfo {author} {\bibfnamefont
  {A.}~\bibnamefont {Riera}}, \bibinfo {author} {\bibfnamefont {M.~N.}\
  \bibnamefont {Bera}},\ and\ \bibinfo {author} {\bibfnamefont
  {M.}~\bibnamefont {Lewenstein}},\ }\bibfield  {title} {\bibinfo {title}
  {{Bounds on the capacity and power of quantum batteries}},\ }\href
  {https://doi.org/10.1103/physrevresearch.2.023113} {\bibfield  {journal}
  {\bibinfo  {journal} {Phys. Rev. Res.}\ }\textbf {\bibinfo {volume} {2}},\
  \bibinfo {pages} {1} (\bibinfo {year} {2020})},\ \Eprint
  {https://arxiv.org/abs/1811.04005} {arXiv:1811.04005} \BibitemShut {NoStop}%
\bibitem [{\citenamefont {Monsel}\ \emph {et~al.}(2020)\citenamefont {Monsel},
  \citenamefont {Fellous-Asiani}, \citenamefont {Huard},\ and\ \citenamefont
  {Auff{\`{e}}ves}}]{Monsel2020}%
  \BibitemOpen
  \bibfield  {author} {\bibinfo {author} {\bibfnamefont {J.}~\bibnamefont
  {Monsel}}, \bibinfo {author} {\bibfnamefont {M.}~\bibnamefont
  {Fellous-Asiani}}, \bibinfo {author} {\bibfnamefont {B.}~\bibnamefont
  {Huard}},\ and\ \bibinfo {author} {\bibfnamefont {A.}~\bibnamefont
  {Auff{\`{e}}ves}},\ }\bibfield  {title} {\bibinfo {title} {{The Energetic
  Cost of Work Extraction}},\ }\href
  {https://doi.org/10.1103/PhysRevLett.124.130601} {\bibfield  {journal}
  {\bibinfo  {journal} {Phys. Rev. Lett.}\ }\textbf {\bibinfo {volume} {124}},\
  \bibinfo {pages} {1} (\bibinfo {year} {2020})},\ \Eprint
  {https://arxiv.org/abs/1907.00812} {arXiv:1907.00812} \BibitemShut {NoStop}%
\bibitem [{\citenamefont {Tirone}\ \emph {et~al.}(2021)\citenamefont {Tirone},
  \citenamefont {Salvia},\ and\ \citenamefont {Giovannetti}}]{Giovannetti2021}%
  \BibitemOpen
  \bibfield  {author} {\bibinfo {author} {\bibfnamefont {S.}~\bibnamefont
  {Tirone}}, \bibinfo {author} {\bibfnamefont {R.}~\bibnamefont {Salvia}},\
  and\ \bibinfo {author} {\bibfnamefont {V.}~\bibnamefont {Giovannetti}},\
  }\bibfield  {title} {\bibinfo {title} {Quantum energy lines and the optimal
  output ergotropy problem},\ }\href
  {https://doi.org/10.1103/PhysRevLett.127.210601} {\bibfield  {journal}
  {\bibinfo  {journal} {Phys. Rev. Lett.}\ }\textbf {\bibinfo {volume} {127}},\
  \bibinfo {pages} {210601} (\bibinfo {year} {2021})}\BibitemShut {NoStop}%
\bibitem [{\citenamefont {Caravelli}\ \emph {et~al.}(2021)\citenamefont
  {Caravelli}, \citenamefont {Yan}, \citenamefont {Garc{\'{i}}a-Pintos},\ and\
  \citenamefont {Hamma}}]{Caravelli2021}%
  \BibitemOpen
  \bibfield  {author} {\bibinfo {author} {\bibfnamefont {F.}~\bibnamefont
  {Caravelli}}, \bibinfo {author} {\bibfnamefont {B.}~\bibnamefont {Yan}},
  \bibinfo {author} {\bibfnamefont {L.~P.}\ \bibnamefont
  {Garc{\'{i}}a-Pintos}},\ and\ \bibinfo {author} {\bibfnamefont
  {A.}~\bibnamefont {Hamma}},\ }\bibfield  {title} {\bibinfo {title} {{Energy
  storage and coherence in closed and open quantum batteries}},\ }\href
  {https://doi.org/10.22331/q-2021-07-15-505} {\bibfield  {journal} {\bibinfo
  {journal} {Quantum}\ }\textbf {\bibinfo {volume} {5}},\ \bibinfo {pages}
  {505} (\bibinfo {year} {2021})},\ \Eprint {https://arxiv.org/abs/2012.15026}
  {arXiv:2012.15026} \BibitemShut {NoStop}%
\bibitem [{\citenamefont {Giazotto}\ \emph {et~al.}(2006)\citenamefont
  {Giazotto}, \citenamefont {Heikkil\"a}, \citenamefont {Luukanen},
  \citenamefont {Savin},\ and\ \citenamefont {Pekola}}]{Pekola_RMP_2006}%
  \BibitemOpen
  \bibfield  {author} {\bibinfo {author} {\bibfnamefont {F.}~\bibnamefont
  {Giazotto}}, \bibinfo {author} {\bibfnamefont {T.~T.}\ \bibnamefont
  {Heikkil\"a}}, \bibinfo {author} {\bibfnamefont {A.}~\bibnamefont
  {Luukanen}}, \bibinfo {author} {\bibfnamefont {A.~M.}\ \bibnamefont
  {Savin}},\ and\ \bibinfo {author} {\bibfnamefont {J.~P.}\ \bibnamefont
  {Pekola}},\ }\bibfield  {title} {\bibinfo {title} {Opportunities for
  mesoscopics in thermometry and refrigeration: Physics and applications},\
  }\href {https://doi.org/10.1103/RevModPhys.78.217} {\bibfield  {journal}
  {\bibinfo  {journal} {Rev. Mod. Phys.}\ }\textbf {\bibinfo {volume} {78}},\
  \bibinfo {pages} {217} (\bibinfo {year} {2006})}\BibitemShut {NoStop}%
\bibitem [{\citenamefont {Sothmann}\ \emph {et~al.}(2014)\citenamefont
  {Sothmann}, \citenamefont {S{\'{a}}nchez},\ and\ \citenamefont
  {Jordan}}]{Jordan_2014}%
  \BibitemOpen
  \bibfield  {author} {\bibinfo {author} {\bibfnamefont {B.}~\bibnamefont
  {Sothmann}}, \bibinfo {author} {\bibfnamefont {R.}~\bibnamefont
  {S{\'{a}}nchez}},\ and\ \bibinfo {author} {\bibfnamefont {A.~N.}\
  \bibnamefont {Jordan}},\ }\bibfield  {title} {\bibinfo {title}
  {Thermoelectric energy harvesting with quantum dots},\ }\href
  {https://doi.org/10.1088/0957-4484/26/3/032001} {\bibfield  {journal}
  {\bibinfo  {journal} {Nanotechnology}\ }\textbf {\bibinfo {volume} {26}},\
  \bibinfo {pages} {032001} (\bibinfo {year} {2014})}\BibitemShut {NoStop}%
\bibitem [{\citenamefont {Pekola}(2015)}]{Pekola_2015}%
  \BibitemOpen
  \bibfield  {author} {\bibinfo {author} {\bibfnamefont {J.}~\bibnamefont
  {Pekola}},\ }\bibfield  {title} {\bibinfo {title} {Towards quantum
  thermodynamics in electronic circuits},\ }\href
  {https://doi.org/10.1038/nphys3169} {\bibfield  {journal} {\bibinfo
  {journal} {Nature Physics}\ }\textbf {\bibinfo {volume} {11}},\ \bibinfo
  {pages} {118} (\bibinfo {year} {2015})}\BibitemShut {NoStop}%
\bibitem [{\citenamefont {Benenti}\ \emph {et~al.}(2017)\citenamefont
  {Benenti}, \citenamefont {Casati}, \citenamefont {Saito},\ and\ \citenamefont
  {Whitney}}]{Whitney2017}%
  \BibitemOpen
  \bibfield  {author} {\bibinfo {author} {\bibfnamefont {G.}~\bibnamefont
  {Benenti}}, \bibinfo {author} {\bibfnamefont {G.}~\bibnamefont {Casati}},
  \bibinfo {author} {\bibfnamefont {K.}~\bibnamefont {Saito}},\ and\ \bibinfo
  {author} {\bibfnamefont {R.~S.}\ \bibnamefont {Whitney}},\ }\bibfield
  {title} {\bibinfo {title} {Fundamental aspects of steady-state conversion of
  heat to work at the nanoscale},\ }\bibfield  {journal} {\bibinfo  {journal}
  {Physics Reports}\ }\textbf {\bibinfo {volume} {694}},\ \href
  {https://doi.org/10.1016/j.physrep.2017.05.008}
  {10.1016/j.physrep.2017.05.008} (\bibinfo {year} {2017})\BibitemShut
  {NoStop}%
\bibitem [{\citenamefont {Freitas}\ and\ \citenamefont
  {Paz}(2017)}]{Freitas2017}%
  \BibitemOpen
  \bibfield  {author} {\bibinfo {author} {\bibfnamefont {N.}~\bibnamefont
  {Freitas}}\ and\ \bibinfo {author} {\bibfnamefont {J.~P.}\ \bibnamefont
  {Paz}},\ }\bibfield  {title} {\bibinfo {title} {Fundamental limits for
  cooling of linear quantum refrigerators},\ }\href
  {https://doi.org/10.1103/PhysRevE.95.012146} {\bibfield  {journal} {\bibinfo
  {journal} {Phys. Rev. E}\ }\textbf {\bibinfo {volume} {95}},\ \bibinfo
  {pages} {012146} (\bibinfo {year} {2017})}\BibitemShut {NoStop}%
\bibitem [{\citenamefont {von Lindenfels}\ \emph {et~al.}(2019)\citenamefont
  {von Lindenfels}, \citenamefont {Gr\"ab}, \citenamefont {Schmiegelow},
  \citenamefont {Kaushal}, \citenamefont {Schulz}, \citenamefont {Mitchison},
  \citenamefont {Goold}, \citenamefont {Schmidt-Kaler},\ and\ \citenamefont
  {Poschinger}}]{Poschinger}%
  \BibitemOpen
  \bibfield  {author} {\bibinfo {author} {\bibfnamefont {D.}~\bibnamefont {von
  Lindenfels}}, \bibinfo {author} {\bibfnamefont {O.}~\bibnamefont {Gr\"ab}},
  \bibinfo {author} {\bibfnamefont {C.~T.}\ \bibnamefont {Schmiegelow}},
  \bibinfo {author} {\bibfnamefont {V.}~\bibnamefont {Kaushal}}, \bibinfo
  {author} {\bibfnamefont {J.}~\bibnamefont {Schulz}}, \bibinfo {author}
  {\bibfnamefont {M.~T.}\ \bibnamefont {Mitchison}}, \bibinfo {author}
  {\bibfnamefont {J.}~\bibnamefont {Goold}}, \bibinfo {author} {\bibfnamefont
  {F.}~\bibnamefont {Schmidt-Kaler}},\ and\ \bibinfo {author} {\bibfnamefont
  {U.~G.}\ \bibnamefont {Poschinger}},\ }\bibfield  {title} {\bibinfo {title}
  {Spin heat engine coupled to a harmonic-oscillator flywheel},\ }\href
  {https://doi.org/10.1103/PhysRevLett.123.080602} {\bibfield  {journal}
  {\bibinfo  {journal} {Phys. Rev. Lett.}\ }\textbf {\bibinfo {volume} {123}},\
  \bibinfo {pages} {080602} (\bibinfo {year} {2019})}\BibitemShut {NoStop}%
\bibitem [{\citenamefont {Maslennikov}\ \emph {et~al.}(2019)\citenamefont
  {Maslennikov}, \citenamefont {Ding}, \citenamefont {Habl{\"u}tzel},
  \citenamefont {Gan}, \citenamefont {Roulet}, \citenamefont {Nimmrichter},
  \citenamefont {Dai}, \citenamefont {Scarani},\ and\ \citenamefont
  {Matsukevich}}]{Matsukevich}%
  \BibitemOpen
  \bibfield  {author} {\bibinfo {author} {\bibfnamefont {G.}~\bibnamefont
  {Maslennikov}}, \bibinfo {author} {\bibfnamefont {S.}~\bibnamefont {Ding}},
  \bibinfo {author} {\bibfnamefont {R.}~\bibnamefont {Habl{\"u}tzel}}, \bibinfo
  {author} {\bibfnamefont {J.}~\bibnamefont {Gan}}, \bibinfo {author}
  {\bibfnamefont {A.}~\bibnamefont {Roulet}}, \bibinfo {author} {\bibfnamefont
  {S.}~\bibnamefont {Nimmrichter}}, \bibinfo {author} {\bibfnamefont
  {J.}~\bibnamefont {Dai}}, \bibinfo {author} {\bibfnamefont {V.}~\bibnamefont
  {Scarani}},\ and\ \bibinfo {author} {\bibfnamefont {D.}~\bibnamefont
  {Matsukevich}},\ }\bibfield  {title} {\bibinfo {title} {Quantum absorption
  refrigerator with trapped ions},\ }\href@noop {} {\bibfield  {journal}
  {\bibinfo  {journal} {Nature communications}\ }\textbf {\bibinfo {volume}
  {10}},\ \bibinfo {pages} {1} (\bibinfo {year} {2019})}\BibitemShut {NoStop}%
\bibitem [{\citenamefont {Ro{\ss}nagel}\ \emph {et~al.}(2016)\citenamefont
  {Ro{\ss}nagel}, \citenamefont {Dawkins}, \citenamefont {Tolazzi},
  \citenamefont {Abah}, \citenamefont {Lutz}, \citenamefont {Schmidt-Kaler},\
  and\ \citenamefont {Singer}}]{Singer}%
  \BibitemOpen
  \bibfield  {author} {\bibinfo {author} {\bibfnamefont {J.}~\bibnamefont
  {Ro{\ss}nagel}}, \bibinfo {author} {\bibfnamefont {S.~T.}\ \bibnamefont
  {Dawkins}}, \bibinfo {author} {\bibfnamefont {K.~N.}\ \bibnamefont
  {Tolazzi}}, \bibinfo {author} {\bibfnamefont {O.}~\bibnamefont {Abah}},
  \bibinfo {author} {\bibfnamefont {E.}~\bibnamefont {Lutz}}, \bibinfo {author}
  {\bibfnamefont {F.}~\bibnamefont {Schmidt-Kaler}},\ and\ \bibinfo {author}
  {\bibfnamefont {K.}~\bibnamefont {Singer}},\ }\bibfield  {title} {\bibinfo
  {title} {A single-atom heat engine},\ }\href@noop {} {\bibfield  {journal}
  {\bibinfo  {journal} {Science}\ }\textbf {\bibinfo {volume} {352}},\ \bibinfo
  {pages} {325} (\bibinfo {year} {2016})}\BibitemShut {NoStop}%
\bibitem [{\citenamefont {Karimi}\ \emph {et~al.}(2020)\citenamefont {Karimi},
  \citenamefont {Brange}, \citenamefont {Samuelsson},\ and\ \citenamefont
  {Pekola}}]{Pekola}%
  \BibitemOpen
  \bibfield  {author} {\bibinfo {author} {\bibfnamefont {B.}~\bibnamefont
  {Karimi}}, \bibinfo {author} {\bibfnamefont {F.}~\bibnamefont {Brange}},
  \bibinfo {author} {\bibfnamefont {P.}~\bibnamefont {Samuelsson}},\ and\
  \bibinfo {author} {\bibfnamefont {J.~P.}\ \bibnamefont {Pekola}},\ }\bibfield
   {title} {\bibinfo {title} {Reaching the ultimate energy resolution of a
  quantum detector},\ }\href@noop {} {\bibfield  {journal} {\bibinfo  {journal}
  {Nature communications}\ }\textbf {\bibinfo {volume} {11}},\ \bibinfo {pages}
  {1} (\bibinfo {year} {2020})}\BibitemShut {NoStop}%
\bibitem [{\citenamefont {Klatzow}\ \emph {et~al.}(2019)\citenamefont
  {Klatzow}, \citenamefont {Becker}, \citenamefont {Ledingham}, \citenamefont
  {Weinzetl}, \citenamefont {Kaczmarek}, \citenamefont {Saunders},
  \citenamefont {Nunn}, \citenamefont {Walmsley}, \citenamefont {Uzdin},\ and\
  \citenamefont {Poem}}]{Walmsley}%
  \BibitemOpen
  \bibfield  {author} {\bibinfo {author} {\bibfnamefont {J.}~\bibnamefont
  {Klatzow}}, \bibinfo {author} {\bibfnamefont {J.~N.}\ \bibnamefont {Becker}},
  \bibinfo {author} {\bibfnamefont {P.~M.}\ \bibnamefont {Ledingham}}, \bibinfo
  {author} {\bibfnamefont {C.}~\bibnamefont {Weinzetl}}, \bibinfo {author}
  {\bibfnamefont {K.~T.}\ \bibnamefont {Kaczmarek}}, \bibinfo {author}
  {\bibfnamefont {D.~J.}\ \bibnamefont {Saunders}}, \bibinfo {author}
  {\bibfnamefont {J.}~\bibnamefont {Nunn}}, \bibinfo {author} {\bibfnamefont
  {I.~A.}\ \bibnamefont {Walmsley}}, \bibinfo {author} {\bibfnamefont
  {R.}~\bibnamefont {Uzdin}},\ and\ \bibinfo {author} {\bibfnamefont
  {E.}~\bibnamefont {Poem}},\ }\bibfield  {title} {\bibinfo {title}
  {Experimental demonstration of quantum effects in the operation of
  microscopic heat engines},\ }\href
  {https://doi.org/10.1103/PhysRevLett.122.110601} {\bibfield  {journal}
  {\bibinfo  {journal} {Phys. Rev. Lett.}\ }\textbf {\bibinfo {volume} {122}},\
  \bibinfo {pages} {110601} (\bibinfo {year} {2019})}\BibitemShut {NoStop}%
\bibitem [{\citenamefont {Hern\'andez-G\'omez}\ \emph {et~al.}(2021)\citenamefont
  {Hernández-Gómez}, \citenamefont {Staudenmaier}, \citenamefont {Campisi},\
  and\ \citenamefont {Fabbri}}]{Campisi21b}%
  \BibitemOpen
  \bibfield  {author} {\bibinfo {author} {\bibfnamefont {S.}~\bibnamefont
  {Hern\'andez-G\'omez}}, \bibinfo {author} {\bibfnamefont {N.}~\bibnamefont
  {Staudenmaier}}, \bibinfo {author} {\bibfnamefont {M.}~\bibnamefont
  {Campisi}},\ and\ \bibinfo {author} {\bibfnamefont {N.}~\bibnamefont
  {Fabbri}},\ }\bibfield  {title} {\bibinfo {title} {Experimental test of
  fluctuation relations for driven open quantum systems with an NV center},\
  }\href {https://doi.org/10.1088/1367-2630/abfc6a} {\bibfield  {journal}
  {\bibinfo  {journal} {New J. Phys.}\ }\textbf {\bibinfo {volume} {23}},\
  \bibinfo {pages} {065004} (\bibinfo {year} {2021})}\BibitemShut {NoStop}%
\bibitem [{\citenamefont {Najera-Santos}\ \emph {et~al.}(2020)\citenamefont
  {Najera-Santos}, \citenamefont {Camati}, \citenamefont {M\'etillon},
  \citenamefont {Brune}, \citenamefont {Raimond}, \citenamefont {Auff\`eves},\
  and\ \citenamefont {Dotsenko}}]{Dotsenko}%
  \BibitemOpen
  \bibfield  {author} {\bibinfo {author} {\bibfnamefont {B.-L.}\ \bibnamefont
  {Najera-Santos}}, \bibinfo {author} {\bibfnamefont {P.~A.}\ \bibnamefont
  {Camati}}, \bibinfo {author} {\bibfnamefont {V.}~\bibnamefont {M\'etillon}},
  \bibinfo {author} {\bibfnamefont {M.}~\bibnamefont {Brune}}, \bibinfo
  {author} {\bibfnamefont {J.-M.}\ \bibnamefont {Raimond}}, \bibinfo {author}
  {\bibfnamefont {A.}~\bibnamefont {Auff\`eves}},\ and\ \bibinfo {author}
  {\bibfnamefont {I.}~\bibnamefont {Dotsenko}},\ }\bibfield  {title} {\bibinfo
  {title} {Autonomous Maxwell's demon in a cavity qed system},\ }\href
  {https://doi.org/10.1103/PhysRevResearch.2.032025} {\bibfield  {journal}
  {\bibinfo  {journal} {Phys. Rev. Research}\ }\textbf {\bibinfo {volume}
  {2}},\ \bibinfo {pages} {032025} (\bibinfo {year} {2020})}\BibitemShut
  {NoStop}%
\bibitem [{\citenamefont {Scigliuzzo}\ \emph {et~al.}(2020)\citenamefont
  {Scigliuzzo}, \citenamefont {Bengtsson}, \citenamefont {Besse}, \citenamefont
  {Wallraff}, \citenamefont {Delsing},\ and\ \citenamefont
  {Gasparinetti}}]{Gasparinetti}%
  \BibitemOpen
  \bibfield  {author} {\bibinfo {author} {\bibfnamefont {M.}~\bibnamefont
  {Scigliuzzo}}, \bibinfo {author} {\bibfnamefont {a.}~\bibnamefont
  {Bengtsson}}, \bibinfo {author} {\bibfnamefont {J.-C.}\ \bibnamefont
  {Besse}}, \bibinfo {author} {\bibfnamefont {a.}~\bibnamefont {Wallraff}},
  \bibinfo {author} {\bibfnamefont {P.}~\bibnamefont {Delsing}},\ and\ \bibinfo
  {author} {\bibfnamefont {S.}~\bibnamefont {Gasparinetti}},\ }\bibfield
  {title} {\bibinfo {title} {Primary thermometry of propagating microwaves in
  the quantum regime},\ }\href {https://doi.org/10.1103/PhysRevX.10.041054}
  {\bibfield  {journal} {\bibinfo  {journal} {Phys. Rev. X}\ }\textbf {\bibinfo
  {volume} {10}},\ \bibinfo {pages} {041054} (\bibinfo {year}
  {2020})}\BibitemShut {NoStop}%
\bibitem [{\citenamefont {Kulikov}\ \emph {et~al.}(2020)\citenamefont
  {Kulikov}, \citenamefont {Navarathna},\ and\ \citenamefont
  {Fedorov}}]{Fedorov}%
  \BibitemOpen
  \bibfield  {author} {\bibinfo {author} {\bibfnamefont {A.}~\bibnamefont
  {Kulikov}}, \bibinfo {author} {\bibfnamefont {R.}~\bibnamefont
  {Navarathna}},\ and\ \bibinfo {author} {\bibfnamefont {A.}~\bibnamefont
  {Fedorov}},\ }\bibfield  {title} {\bibinfo {title} {Measuring effective
  temperatures of qubits using correlations},\ }\href
  {https://doi.org/10.1103/PhysRevLett.124.240501} {\bibfield  {journal}
  {\bibinfo  {journal} {Phys. Rev. Lett.}\ }\textbf {\bibinfo {volume} {124}},\
  \bibinfo {pages} {240501} (\bibinfo {year} {2020})}\BibitemShut {NoStop}%
\bibitem [{\citenamefont {Peterson}\ \emph {et~al.}(2019)\citenamefont
  {Peterson}, \citenamefont {Batalh{\~a}o}, \citenamefont {Herrera},
  \citenamefont {Souza}, \citenamefont {Sarthour}, \citenamefont {Oliveira},\
  and\ \citenamefont {Serra}}]{Serra}%
  \BibitemOpen
  \bibfield  {author} {\bibinfo {author} {\bibfnamefont {J.~P.}\ \bibnamefont
  {Peterson}}, \bibinfo {author} {\bibfnamefont {T.~B.}\ \bibnamefont
  {Batalh{\~a}o}}, \bibinfo {author} {\bibfnamefont {M.}~\bibnamefont
  {Herrera}}, \bibinfo {author} {\bibfnamefont {A.~M.}\ \bibnamefont {Souza}},
  \bibinfo {author} {\bibfnamefont {R.~S.}\ \bibnamefont {Sarthour}}, \bibinfo
  {author} {\bibfnamefont {I.~S.}\ \bibnamefont {Oliveira}},\ and\ \bibinfo
  {author} {\bibfnamefont {R.~M.}\ \bibnamefont {Serra}},\ }\bibfield  {title}
  {\bibinfo {title} {Experimental characterization of a spin quantum heat
  engine},\ }\href@noop {} {\bibfield  {journal} {\bibinfo  {journal} {Physical
  Review Letters}\ }\textbf {\bibinfo {volume} {123}},\ \bibinfo {pages}
  {240601} (\bibinfo {year} {2019})}\BibitemShut {NoStop}%
\bibitem [{\citenamefont {Gluza}\ \emph {et~al.}(2021)\citenamefont {Gluza},
  \citenamefont {Sabino}, \citenamefont {Ng}, \citenamefont {Vitagliano},
  \citenamefont {Pezzutto}, \citenamefont {Omar}, \citenamefont {Mazets},
  \citenamefont {Huber}, \citenamefont {Schmiedmayer},\ and\ \citenamefont
  {Eisert}}]{Schmiedmayer}%
  \BibitemOpen
  \bibfield  {author} {\bibinfo {author} {\bibfnamefont {M.}~\bibnamefont
  {Gluza}}, \bibinfo {author} {\bibfnamefont {J.~A.}\ \bibnamefont {Sabino}},
  \bibinfo {author} {\bibfnamefont {N.~H.}\ \bibnamefont {Ng}}, \bibinfo
  {author} {\bibfnamefont {G.}~\bibnamefont {Vitagliano}}, \bibinfo {author}
  {\bibfnamefont {M.}~\bibnamefont {Pezzutto}}, \bibinfo {author}
  {\bibfnamefont {Y.}~\bibnamefont {Omar}}, \bibinfo {author} {\bibfnamefont
  {I.}~\bibnamefont {Mazets}}, \bibinfo {author} {\bibfnamefont
  {M.}~\bibnamefont {Huber}}, \bibinfo {author} {\bibfnamefont
  {J.}~\bibnamefont {Schmiedmayer}},\ and\ \bibinfo {author} {\bibfnamefont
  {J.}~\bibnamefont {Eisert}},\ }\bibfield  {title} {\bibinfo {title} {Quantum
  field thermal machines},\ }\href
  {https://doi.org/10.1103/PRXQuantum.2.030310} {\bibfield  {journal} {\bibinfo
   {journal} {PRX Quantum}\ }\textbf {\bibinfo {volume} {2}},\ \bibinfo {pages}
  {030310} (\bibinfo {year} {2021})}\BibitemShut {NoStop}%
\bibitem [{\citenamefont {Brunelli}\ \emph {et~al.}(2018)\citenamefont
  {Brunelli}, \citenamefont {Fusco}, \citenamefont {Landig}, \citenamefont
  {Wieczorek}, \citenamefont {Hoelscher-Obermaier}, \citenamefont {Landi},
  \citenamefont {Semi\~ao}, \citenamefont {Ferraro}, \citenamefont {Kiesel},
  \citenamefont {Donner}, \citenamefont {De~Chiara},\ and\ \citenamefont
  {Paternostro}}]{Donner}%
  \BibitemOpen
  \bibfield  {author} {\bibinfo {author} {\bibfnamefont {M.}~\bibnamefont
  {Brunelli}}, \bibinfo {author} {\bibfnamefont {L.}~\bibnamefont {Fusco}},
  \bibinfo {author} {\bibfnamefont {R.}~\bibnamefont {Landig}}, \bibinfo
  {author} {\bibfnamefont {W.}~\bibnamefont {Wieczorek}}, \bibinfo {author}
  {\bibfnamefont {J.}~\bibnamefont {Hoelscher-Obermaier}}, \bibinfo {author}
  {\bibfnamefont {G.}~\bibnamefont {Landi}}, \bibinfo {author} {\bibfnamefont
  {F.~L.}\ \bibnamefont {Semi\~ao}}, \bibinfo {author} {\bibfnamefont
  {A.}~\bibnamefont {Ferraro}}, \bibinfo {author} {\bibfnamefont
  {N.}~\bibnamefont {Kiesel}}, \bibinfo {author} {\bibfnamefont
  {T.}~\bibnamefont {Donner}}, \bibinfo {author} {\bibfnamefont
  {G.}~\bibnamefont {De~Chiara}},\ and\ \bibinfo {author} {\bibfnamefont
  {M.}~\bibnamefont {Paternostro}},\ }\bibfield  {title} {\bibinfo {title}
  {Experimental determination of irreversible entropy production in
  out-of-equilibrium mesoscopic quantum systems},\ }\href
  {https://doi.org/10.1103/PhysRevLett.121.160604} {\bibfield  {journal}
  {\bibinfo  {journal} {Phys. Rev. Lett.}\ }\textbf {\bibinfo {volume} {121}},\
  \bibinfo {pages} {160604} (\bibinfo {year} {2018})}\BibitemShut {NoStop}%
\bibitem [{\citenamefont {Debiossac}\ \emph {et~al.}(2020)\citenamefont
  {Debiossac}, \citenamefont {Grass}, \citenamefont {Alonso}, \citenamefont
  {Lutz},\ and\ \citenamefont {Kiesel}}]{Kiesel}%
  \BibitemOpen
  \bibfield  {author} {\bibinfo {author} {\bibfnamefont {M.}~\bibnamefont
  {Debiossac}}, \bibinfo {author} {\bibfnamefont {D.}~\bibnamefont {Grass}},
  \bibinfo {author} {\bibfnamefont {J.}~\bibnamefont {Alonso}}, \bibinfo
  {author} {\bibfnamefont {E.}~\bibnamefont {Lutz}},\ and\ \bibinfo {author}
  {\bibfnamefont {N.}~\bibnamefont {Kiesel}},\ }\bibfield  {title} {\bibinfo
  {title} {Thermodynamics of continuous non-markovian feedback control},\
  }\href {https://doi.org/10.1038/s41467-020-15148-5} {\bibfield  {journal}
  {\bibinfo  {journal} {Nat Commun}\ }\textbf {\bibinfo {volume} {11}},\
  \bibinfo {pages} {1330} (\bibinfo {year} {2020})}\BibitemShut {NoStop}%
\bibitem [{\citenamefont {Pearson}\ \emph {et~al.}(2021)\citenamefont
  {Pearson}, \citenamefont {Guryanova}, \citenamefont {Erker}, \citenamefont
  {Laird}, \citenamefont {Briggs}, \citenamefont {Huber},\ and\ \citenamefont
  {Ares}}]{Ares}%
  \BibitemOpen
  \bibfield  {author} {\bibinfo {author} {\bibfnamefont {A.~N.}\ \bibnamefont
  {Pearson}}, \bibinfo {author} {\bibfnamefont {Y.}~\bibnamefont {Guryanova}},
  \bibinfo {author} {\bibfnamefont {P.}~\bibnamefont {Erker}}, \bibinfo
  {author} {\bibfnamefont {E.~A.}\ \bibnamefont {Laird}}, \bibinfo {author}
  {\bibfnamefont {G.~A.~D.}\ \bibnamefont {Briggs}}, \bibinfo {author}
  {\bibfnamefont {M.}~\bibnamefont {Huber}},\ and\ \bibinfo {author}
  {\bibfnamefont {N.}~\bibnamefont {Ares}},\ }\bibfield  {title} {\bibinfo
  {title} {Measuring the thermodynamic cost of timekeeping},\ }\href
  {https://doi.org/10.1103/PhysRevX.11.021029} {\bibfield  {journal} {\bibinfo
  {journal} {Phys. Rev. X}\ }\textbf {\bibinfo {volume} {11}},\ \bibinfo
  {pages} {021029} (\bibinfo {year} {2021})}\BibitemShut {NoStop}%
\bibitem [{\citenamefont {Solfanelli}\ \emph {et~al.}(2021)\citenamefont
  {Solfanelli}, \citenamefont {Santini},\ and\ \citenamefont
  {Campisi}}]{Campisi21c}%
  \BibitemOpen
  \bibfield  {author} {\bibinfo {author} {\bibfnamefont {A.}~\bibnamefont
  {Solfanelli}}, \bibinfo {author} {\bibfnamefont {A.}~\bibnamefont
  {Santini}},\ and\ \bibinfo {author} {\bibfnamefont {M.}~\bibnamefont
  {Campisi}},\ }\bibfield  {title} {\bibinfo {title} {Experimental verification
  of fluctuation relations with a quantum computer},\ }\href
  {https://doi.org/10.1103/PRXQuantum.2.030353} {\bibfield  {journal} {\bibinfo
   {journal} {PRX Quantum}\ }\textbf {\bibinfo {volume} {2}},\ \bibinfo {pages}
  {030353} (\bibinfo {year} {2021})}\BibitemShut {NoStop}%
\bibitem [{\citenamefont {Fellous-Asiani}\ \emph {et~al.}(2021)\citenamefont
  {Fellous-Asiani}, \citenamefont {Chai}, \citenamefont {Whitney},
  \citenamefont {Auff\`eves},\ and\ \citenamefont {Ng}}]{Marco_2021}%
  \BibitemOpen
  \bibfield  {author} {\bibinfo {author} {\bibfnamefont {M.}~\bibnamefont
  {Fellous-Asiani}}, \bibinfo {author} {\bibfnamefont {J.~H.}\ \bibnamefont
  {Chai}}, \bibinfo {author} {\bibfnamefont {R.~S.}\ \bibnamefont {Whitney}},
  \bibinfo {author} {\bibfnamefont {A.}~\bibnamefont {Auff\`eves}},\ and\
  \bibinfo {author} {\bibfnamefont {H.~K.}\ \bibnamefont {Ng}},\ }\bibfield
  {title} {\bibinfo {title} {Limitations in quantum computing from resource
  constraints},\ }\href {https://doi.org/10.1103/PRXQuantum.2.040335}
  {\bibfield  {journal} {\bibinfo  {journal} {PRX Quantum}\ }\textbf {\bibinfo
  {volume} {2}},\ \bibinfo {pages} {040335} (\bibinfo {year}
  {2021})}\BibitemShut {NoStop}%
\bibitem [{\citenamefont {Fellous-Asiani}\ \emph {et~al.}()\citenamefont
  {Fellous-Asiani}, \citenamefont {Chai}, \citenamefont {Ng}, \citenamefont
  {Whitney},\ and\ \citenamefont {Auff\`eves}}]{Marco_2022}%
  \BibitemOpen
  \bibfield  {author} {\bibinfo {author} {\bibfnamefont {M.}~\bibnamefont
  {Fellous-Asiani}}, \bibinfo {author} {\bibfnamefont {J.~H.}\ \bibnamefont
  {Chai}}, \bibinfo {author} {\bibfnamefont {H.~K.}\ \bibnamefont {Ng}},
  \bibinfo {author} {\bibfnamefont {R.~S.}\ \bibnamefont {Whitney}},\ and\
  \bibinfo {author} {\bibfnamefont {A.}~\bibnamefont {Auff\`eves}},\ }\bibfield
   {title} {\bibinfo {title} {Optimizing energetic efficiency for scalable
  full-stack quantum computers},\ }\href@noop {} {\bibinfo  {journal} {in
  prep}\ }\BibitemShut {NoStop}%
\bibitem [{\citenamefont {Fellous-Asiani}()}]{Marco_PhD}%
  \BibitemOpen
\bibfield  {journal} {  }\bibfield  {author} {\bibinfo {author} {\bibfnamefont
  {M.}~\bibnamefont {Fellous-Asiani}},\ }\bibfield  {title} {\bibinfo {title}
  {The resource cost of large scale quantum computing},\ }\bibfield  {journal}
  {\bibinfo  {journal} {PhD manuscript}\ }\href
  {https://doi.org/https://doi.org/10.48550/arXiv.2112.04022}
  {https://doi.org/10.48550/arXiv.2112.04022}\BibitemShut {NoStop}%
\bibitem [{\citenamefont {Gidney}\ and\ \citenamefont
  {Eker{\aa{}}}(2021)}]{Gidney}%
  \BibitemOpen
  \bibfield  {author} {\bibinfo {author} {\bibfnamefont {C.}~\bibnamefont
  {Gidney}}\ and\ \bibinfo {author} {\bibfnamefont {M.}~\bibnamefont
  {Eker{\aa{}}}},\ }\bibfield  {title} {\bibinfo {title} {How to factor 2048
  bit {RSA} integers in 8 hours using 20 million noisy qubits},\ }\href
  {https://doi.org/10.22331/q-2021-04-15-433} {\bibfield  {journal} {\bibinfo
  {journal} {{Quantum}}\ }\textbf {\bibinfo {volume} {5}},\ \bibinfo {pages}
  {433} (\bibinfo {year} {2021})}\BibitemShut {NoStop}%
\bibitem [{\citenamefont {Zhou}\ \emph {et~al.}(2020)\citenamefont {Zhou},
  \citenamefont {Stoudenmire},\ and\ \citenamefont {Waintal}}]{Waintal}%
  \BibitemOpen
  \bibfield  {author} {\bibinfo {author} {\bibfnamefont {Y.}~\bibnamefont
  {Zhou}}, \bibinfo {author} {\bibfnamefont {E.~M.}\ \bibnamefont
  {Stoudenmire}},\ and\ \bibinfo {author} {\bibfnamefont {X.}~\bibnamefont
  {Waintal}},\ }\bibfield  {title} {\bibinfo {title} {What limits the
  simulation of quantum computers?},\ }\href
  {https://doi.org/10.1103/PhysRevX.10.041038} {\bibfield  {journal} {\bibinfo
  {journal} {Phys. Rev. X}\ }\textbf {\bibinfo {volume} {10}},\ \bibinfo
  {pages} {041038} (\bibinfo {year} {2020})}\BibitemShut {NoStop}%
\bibitem [{\citenamefont {Berger}\ \emph {et~al.}()\citenamefont {Berger},
  \citenamefont {Di~Paolo}, \citenamefont {Forrest}, \citenamefont {Hadfield},
  \citenamefont {Sawaya}, \citenamefont {Stechly},\ and\ \citenamefont
  {Thibault}}]{Qforclimate}%
  \BibitemOpen
  \bibfield  {author} {\bibinfo {author} {\bibfnamefont {C.}~\bibnamefont
  {Berger}}, \bibinfo {author} {\bibfnamefont {A.}~\bibnamefont {Di~Paolo}},
  \bibinfo {author} {\bibfnamefont {T.}~\bibnamefont {Forrest}}, \bibinfo
  {author} {\bibfnamefont {S.}~\bibnamefont {Hadfield}}, \bibinfo {author}
  {\bibfnamefont {N.}~\bibnamefont {Sawaya}}, \bibinfo {author} {\bibfnamefont
  {M.}~\bibnamefont {Stechly}},\ and\ \bibinfo {author} {\bibfnamefont
  {K.}~\bibnamefont {Thibault}},\ }\bibfield  {title} {\bibinfo {title}
  {Quantum technologies for climate change: Preliminary assessment},\
  }\href@noop {} {\bibfield  {journal} {\bibinfo  {journal} {arXiv}\ }\textbf
  {\bibinfo {volume} {2107.05362}}}\BibitemShut {NoStop}%
\bibitem [{\citenamefont {Ezratty}()}]{hype}%
  \BibitemOpen
  \bibfield  {author} {\bibinfo {author} {\bibfnamefont {O.}~\bibnamefont
  {Ezratty}},\ }\href {https://doi.org/10.48550/arXiv.2202.01925} {\bibfield
  {journal} {\bibinfo  {journal} {arXiv}\ }\textbf {\bibinfo {volume}
  {2202.01925}}}\BibitemShut {NoStop}%
\end{thebibliography}
\end{document}